\documentclass{ws-ijmpd}
\usepackage[super,compress]{cite}
\usepackage{color}
\usepackage{verbatim}
\usepackage{textcomp}
\usepackage{amsmath}
\usepackage{amssymb}
\usepackage{graphicx}
\usepackage{esint}
\usepackage{ulem}
\usepackage{bm}

\begin{document}

\markboth{Georgios Lukes-Gerakopoulos, Ond\v{r}ej Kop\'{a}\v{c}ek}
{Recurrence Analysis as a tool to study chaotic dynamics of extreme mass ratio inspiral in signal with noise}

%
\catchline{}{}{}{}{}
%

\title{Recurrence Analysis as a tool to study chaotic dynamics of extreme mass ratio inspiral in signal with noise}

\author{Georgios Lukes-Gerakopoulos}
\address{Astronomical Institute of the Academy of Sciences of the Czech Republic,
Bo\v{c}n\'{i} II 1401/1a,\\
CZ-141 31 Prague, Czech Republic\\
gglukes@gmail.com}
\author{Ond\v{r}ej Kop\'{a}\v{c}ek}
\address{Astronomical Institute of the Academy of Sciences of the Czech Republic,
Bo\v{c}n\'{i} II 1401/1a, CZ-141 31 Prague, Czech Republic\\
kopacek@ig.cas.cz}

\maketitle

\begin{history}
\received{Day Month Year}
\revised{Day Month Year}
\end{history}


%

%

%

\begin{abstract}
 Recurrence analysis is a well settled method allowing to discern chaos from order,
 and determinism from noise. We apply this tool to study time series representing
 geodesic and inspiraling motion of a test particle in a deformed Kerr 
 spacetime, when deterministic chaos and different levels of stochastic
 noise are present. In particular, we suggest a recurrence-based criterion to
 reveal whether the time series comes from a deterministic source and find a 
 noise-level threshold of its applicability.
\end{abstract}

\keywords{black holes;chaos}

\ccode{04.25.dg; 04.70.Bw; 95.10.Fh}

 \section{Introduction}

 Extreme mass ratio inspirals (EMRIs) are highly promising sources of
 gravitational waves for detectors like the Laser Interferometer Space Antenna
 (LISA) \cite{LISA}. To form an EMRI, a stellar compact object, e.g., a black
 hole or a neutron star, has to be trapped in the gravitational field of a 
 supermassive black hole (SMBH). As the stellar compact object is orbiting
 the SMBH, gravitational waves are emitted forcing the stellar object to follow
 an inspiraling trajectory. To detect these gravitational 
 waves, template banks are built in order to use them for the matched filtering
 method. For constructing these templates it is assumed that the spacetime
 background is described by the Kerr metric and that the inspiraling compact
 object shifts adiabatically from a geodesic orbit to a geodesic orbit
 \cite{wavetemp}. The above standard {\it adiabatic approximation} implies that
 the geodesic orbits are regular since Kerr spacetime corresponds to an
 integrable system. If the system is not integrable, then non-linear phenomena
 like chaos and prolonged resonances will appear \cite{Kiuchi04,nonKerr}.
 However, the current template banks do not include such scenarios, which means
 that in such cases matched filtering would fail.
  
 In fact, the standard adiabatic approximation neglects various factors leading
 to non-integrability. For example, if the spin of the inspiraling object is taken
 into account then chaotic motion appears \cite{Suzuki97,Hartl03,Hartl03b,Han08a};
 the presence of an outer stellar object orbiting around an EMRI can cause
 ``butterfly'' effects to the inspiraling object \cite{AmaroSeoane12}; the
 presence of a matter distribution, like a halo or a ring, around a black
 hole acts as a source of chaos \cite{Vieira99,Wu06,SemSuk10,SemSuk12,SemSuk13,SemSuk15}.
 In the above cases the Kerr black hole hypothesis \cite{Bambi11} is assumed to
 hold for the SMBH. However, it has been shown that spacetimes diverging from
 the Kerr by a deformation parameter can also result in non-integrable systems
 \cite{nonKerr,Gueron,Gair08,Han08b}. The apparent question is how different 
 would be the gravitational waveforms coming from chaotic orbits compared
 with those coming from regular ones.
 
To address this question the emitted gravitational radiation
 of spinning objects in a Kerr background was studied by Kiuchi and Maeda \cite{Kiuchi04}. First they investigated
 whether the waveforms resulting from a regular and a chaotic orbit bear any 
 significant differences and they found that the respective waveforms look similar.
 Then they applied a Fourier transform on the waveforms to obtain the corresponding
 energy spectra. In the case of a chaotic orbit they found a continuous spectrum
 with several peaks, while for a regular orbit the spectrum was discrete. This is
 a basic method to distinguish between chaotic and regular orbits in a conservative
 system, see, e.g., Ref.~\refcite{Contop02}. However, in an actual EMRI the dissipation
 due to the radiation reaction is always present and the inspiraling object is
 losing energy and angular momentum. By including the dissipation factor, the
 peaks of the spectrum shift, since the trajectory itself shifts from an orbit
 to an orbit. The result of such a shift is that for a regular orbit one does
 not necessarily obtain a discrete spectrum anymore and in this sense the
 spectra of the regular and the chaotic orbit become similar. In fact, for weak
 chaos in conservative systems, such shifts of peaks are used as an indicator to show 
 that an orbit is chaotic \cite{Laskar93}. The problem of detecting gravitational
 waves coming from potentially chaotic orbits is even more complicated since the
 received waveforms are expected to be buried in various types of noise
 \cite{LISA}. Thus, the Fourier spectrum of a  real signal is always contaminated
 by noise and the identification of the deterministic signal becomes intricate 
 even for the conservative systems.
 
 Since the existing template banks do not include scenarios where non-linear
 phenomena are present in the dynamics of the source, the question is how one
 can distinguish noise from deterministic signal without selecting a particular
 model of an EMRI. 
 Although, the current work does not answer the given question, it provides a 
 study of a particular statistical method that is applied to simulated data from
 a dynamical system that is a simplified model of EMRI. Namely, we use a
 statistical analysis of recurrences occurring in time series produced by a body
 moving in Manko, Sanabria-G\'{o}mez, Manko (MSM) spacetime \cite{MSM}, since 
 MSM provides chaotic geodesic orbits \cite{Dubeibe07,Han08b,geochaos}, to study
 up to which noise level geodesic chaos can be detected even when dissipation is
 present. The MSM spacetime is an exact vacuum solution of the Einstein's field
 equations and describes the ``exterior field of a charged, magnetized, spinning
 deformed mass''\cite{MSM}. The MSM spacetime serves in our analysis as an ``easy'' way to 
 simulate a non-integrable EMRI system, and this choice does not restrict the
 generality of our analysis. This analysis could be applied with any of the above
 mentioned parameters inducing a non-integrable behavior of an EMRI, e.g., by 
 including the spin of the particle just like Kiuchi and Maeda did \cite{Kiuchi04}.
 This is something we actually plan to do once the gravitational wave fluxes
 from spinning particles are obtained, see, e.g., Ref.~\refcite{SPGW}.

 The rest of the article is organized as follows. In Sec.~\ref{sec:MSM} the MSM
 spacetime is introduced. Sec.~\ref{sec:Chaos} discusses the Lyapunov number and
 the recurrence analysis. In Sec.~\ref{sec:RQA} the recurrence analysis is 
 applied on simulated data. At first we inspect the case of geodesic motion in a
 MSM spacetime with different noise levels, and then we also consider the 
 generalized scenario including the dissipation due to radiation reaction.
 Sec.~\ref{sec:Conc} concludes the paper. The appendix provides
 technical details about the parameters used in the recurrence analysis.

 \section{Spacetime background} \label{sec:MSM}
 
 The MSM metric was designed to model neutron stars \cite{MSM,Berti04,Berti05},
 but here we reduce it to a deformed Kerr spacetime which is known to trigger
 geodesic chaos \cite{Dubeibe07,Han08b}. The MSM spacetime depends on five real
 parameters describing this central object: the mass $M$,
 the spin $a$ (per unit mass $M$), the total charge $q$, the magnetic dipole
 moment ${\cal M}$, and the mass-quadrupole moment ${\cal Q}$. The latter two
 parameters are actually represented in the metric as functions of two other
 parameters $b$ and $\mu$, i.e.
 \begin{align} 
  {\cal M} &= \mu+q (a-b) \quad, \label{eq:MagDip} \\
  {\cal Q} &= -M(d-\delta-a~b+a^2) \quad, \label{eq:MassQuad}
 \end{align}
 where 
 \begin{align} \label{eq:deltad}
   \delta := \frac{\mu^2-M^2 b^2}{M^2-(a-b)^2-q^2}, \quad
   d := \frac{1}{4}[M^2-(a-b)^2-q^2] \quad.  
 \end{align}
 For the purpose of our work we set the charge $q$ and the parameter $\mu$ to
 zero, thus, the magnetic dipole moment~\eqref{eq:MagDip} is set to zero.
 This also reduces the mass-quadrupole moment~\eqref{eq:MassQuad} to
 \begin{align} \label{eq:MassQuadRed}
  {\cal Q}=-M \left(\frac{(M^2-(a-b)^2)^2+4~M^2~b^2}{4(M^2-(a-b)^2)}-a~b+a^2\right) \quad,
 \end{align}
 where $b$ is now the only parameter making the MSM spacetime to deviate from a
 Kerr one. In particular, when $b^2=a^2-M^2$ Eq.~\eqref{eq:MassQuadRed} gives
 ${\cal Q}_\textrm{Kerr}=-a M^2$, and the MSM spacetime reduces to Kerr.
 If we define the quadrupole deviation parameter
 \begin{align}\label{eq:QuadDeform}
  q:={\cal Q}-{\cal Q}_\textrm{Kerr}\quad,
 \end{align}
 then for $q>0$ the MSM describes a more
 prolate compact object than the Kerr black hole and for $q<0$ a more oblate
 one, while for $q=0$ the MSM identifies with the Kerr black hole. Thus,
 in this sense MSM is a bumpy black hole, i.e., a compact object deviating from 
 a Kerr black hole\cite{Gair08,nonKerr,Bambi11}. Such bumpy black holes are 
 often employed to test the Kerr hypothesis in EMRIS\cite{Gair08,nonKerr,Bambi11}.

 A line element in Weyl-Papapetrou coordinates reads
 \begin{align}\label{eq:MNSpro}
  ds^2  =  -f(dt-\omega d\phi)^2 
        +  f^{-1} \left[ e^{2\gamma} (d\rho^2+dz^2)+\rho^2 d\phi^2 \right]\quad.
 \end{align}
 For the MSM the metric functions read  
 \begin{align} \label{eq:MetrFunc}
  f = {\cal E}/D,\quad 
  e^{2\gamma} = {\cal E}/16 \kappa^8 (u^2-v^2)^4,\quad 
  \omega = (v^2-1)F/{\cal E} \quad,
 \end{align}
 where
 \begin{align} \label{eq:EDF}
  {\cal E} := R^2+\lambda_1\lambda_2 S^2,~
  D :=  {\cal E}+R P+\lambda_2 S T,~
  F := R T-\lambda_1 S P ~~,
 \end{align}
 \begin{align} \label{eq:lambda}
  \lambda_1 :=\kappa^2 (u^2-1),\quad \lambda_2 :=v^2-1 \quad,
 \end{align}
 and

 \begin{align}\label{eq:PRST}
  P & :=  2 \{\kappa M u [(2 \kappa u+M)^2-2 v^2 (2 \delta +a b -b^2) \nonumber \\
    &  -  a^2+b^2 -q^2]-2 \kappa^2 q^2 u^2-2 v^2 (4 \delta d-M^2 b^2)\}\quad, 
   \nonumber \\
  R & :=  4 [\kappa^2 (u^2-1)+\delta (1-v^2)]^2 
  +  (a-b) [(a-b)(d-\delta)-M^2 b +q~\mu](1-v^2)^2\quad,\nonumber \\
  S & :=  -4 {(a-b)[\kappa^2(u^2-v^2)+2 \delta v^2]+v^2 (M^2 b-q~\mu)}\quad,
  \nonumber \\
  T & :=  4(2 \kappa M b u+2 M^2 b-q~\mu)[\kappa^2 (u^2-1)+\delta (1-v^2)]
     +  (1-v^2)\{(a-b)(M^2 b^2 -4 \delta d)  \nonumber \\
    & -  (4 \kappa M u+2 M^2-q^2) [(a-b)(d-\delta)-M^2 b +q~\mu]\}\quad.
 \end{align}
 
 All the metric functions are expressed in prolate spheroidal coordinates $u,~v$.
 However, the line element~\eqref{eq:MNSpro} is written in the corresponding
 cylindrical coordinates $\rho,~z$. The transformation between these coordinate
 systems is given by
 \begin{align} \label{eq:TrCylPS}
  \rho &=\kappa \sqrt{(u^2-1)(1-v^2)},\quad z=\kappa u v\quad,
 \end{align}

 where
 \begin{align} \label{eq:kappa}
  \kappa :=\sqrt{d+\delta}\quad. 
 \end{align}
 
 \subsection{Geodesic motion} \label{subsec-II-b}

 The Lagrangian function
 \begin{align} \label{eq:LagDef}
  L=\frac{1}{2} g_{\mu\nu}~ \dot{x}^{\mu} \dot{x}^{\nu}\quad
 \end{align}
 provides the equations of geodesic motion in a spacetime with metric $g_{\mu\nu}$. The dot
 denotes derivation with respect to proper time $\tau$, thus the
 Lagrangian~\eqref{eq:LagDef} expresses the four-velocity
 $g_{\mu\nu}~\dot{x}^{\mu}\dot{x}^{\nu}=-1$ constraint.

 Since the spacetime is stationary and axisymmetric, the specific energy
 \begin{align}  \label{eq:EnCon}
  E=-\frac{\partial L}{\partial \dot{t} }\quad,
 \end{align}
 and the specific azimuthal component of the angular momentum
 \begin{align} \label{eq:AnMomCon}
  L_z =\frac{\partial L}{\partial \dot{\phi}}\quad,
 \end{align}
 are conserved. For simplicity, we set $M=1$, which is 
 equivalent with replacing all the quantities with their dimensionless   
 counterparts, e.g., $\tau/M$, $\rho/M$. Thus, we refer hereafter to the above
 two integrals of motion as the energy and the angular momentum. 
 
 Through Eqs.~\eqref{eq:EnCon} and \eqref{eq:AnMomCon} we restrict the motion on
 the meridian plane. Thus, from the original set of $4$ coupled second order
 ordinary differential equations, we arrive to a set of $2$ coupled ODEs. 

 \section{Chaos Detection} \label{sec:Chaos}

 \subsection{Lyapunov number} \label{sec:LyapNum}

 A common way to distinguish between chaotic and regular orbits are indicators
 depending on the deviation vector $\xi^\alpha$. A deviation vector can be 
 interpreted as an indicator of how two initially nearby worldlines $x^\alpha$ and 
 $x^\alpha+\xi^\alpha$ diverge from each other. The evolution of the deviation
 vector along the geodesic orbit is provided by the geodesic deviation equation
 \begin{align} \label{eq:GeoDev}
  \ddot{\xi}^\alpha+2\Gamma^\alpha_{\beta\gamma}\dot{x}^\beta\dot{\xi}^\gamma+
 \frac{\partial \Gamma^\alpha_{\beta\gamma}}{\partial x^\delta}
 \dot{x}^\beta\dot{x}^\gamma\xi^\delta=0~~.
 \end{align}
 However, a deviation vector in the classical framework should represent the
 phase space neighborhood around a point of the trajectory and not only the
 configuration space. A deviation vector should show how the phase space is
 stretched and folded in the neighborhood of the trajectory during the evolution.
 To address this, Sota et al. in Ref.~\refcite{Sota96} defined an invariant
 measure of the deviation vector as
 \begin{align} \label{eq:SotaDis}
   \Xi^2=g_{\alpha\beta}\xi^\alpha \xi^\beta+g_{\alpha\beta}
 \frac{D\xi^\alpha}{d\tau}\frac{D\xi^\beta}{d\tau}~~,
 \end{align}
 where the covariant derivative
 \begin{align}\label{eq:CovDev}
   \frac{D\xi^\alpha}{d\tau}=\dot{\xi}^\alpha
   +\Gamma^\alpha_{\beta\gamma} \dot{x}^\beta \xi^\gamma
 \end{align}
 provides the divergence of the velocities.
 
 For a regular orbit the measure of the deviation vector grows linearly, while
 for a chaotic one, it grows exponentially, see, e.g., Ref.~\refcite{Skokos10}. As an
 invariant measure of the evolution parameter the proper time normalized by the 
 time scale $G M/c^3$ was used in Ref.~\refcite{Sota96}, which reduces to $M$ 
 in our case, as we employ geometric units $(G=c=1)$.

 To measure chaos, usually the maximal Lyapunov Characteristic Exponent 
 \begin{align}\label{eq:mLCE}
  \displaystyle \textrm{mLCE}=\frac{1}{\tau}\ln{\frac{\Xi(\tau)}{\Xi(0)}}
 \end{align}
 is utilized. Even if in theory the mLCE should be calculated for the
 $\tau\rightarrow\infty$ limit, practically $\tau$ has a large but finite value.
 Lyapunov numbers are a standard way to measure chaos, but they cannot distinguish 
 chaos from noise. To address the latter issue  we employ methods based on the
 statistical analysis of recurrences in a data time series, i.e., the recurrence 
 plots and recurrence quantification analysis \cite{marwan11}.
 
 \subsection{Recurrence Analysis} \label{sec:RecAna}
   
 The recurrence of a general (vector) time series $\bm{y}(t)$ occurs when the
 distance (measured in an abstract phase space) between the $i$th point and the
 $j$th point of the data series drops below a pre-defined threshold $\varepsilon$.
 Recurrences are recorded in the (binary) recurrence matrix 
 \begin{align} \label{eq:RecMat}
   \mathbf{R}_{ij}(\varepsilon)=\Theta(\varepsilon-||\bm{y}(i)-\bm{y}(j)||)\quad,
 \end{align}
 where $||.||$ is the norm, and $\Theta$ stands for the Heaviside step-function.
 Recurrence matrix $\mathbf{R}_{ij}$ may be visualized as a recurrence plot (RP)
 \cite{eckmann87}. Actually, RPs encode surprising amount of fundamental information
 about the dynamics of the trajectory \cite{thiel04}. Just a visual survey
 of RPs provides an intuitive method of how to discern regular from chaotic motion
 and deterministic signal from noise \cite{thiel02,marwan07}. An optimal choice of
 the recurrence threshold $\varepsilon$ is crucial for the reliable outcome of
 the analysis. Several rules of thumb have been suggested, e.g., to use the
 value of $\varepsilon$ which corresponds to 10\% of the mean phase space diameter,
 or 25\% of the standard deviation of the analyzed data \cite{schinkel08}.
 Nevertheless, for our application the optimal method is to set $\varepsilon$
 which ensures given recurrences point density (namely 2\% in our case).
 
  \begin{figure}
  \centerline{ \includegraphics[width=0.45\textwidth]{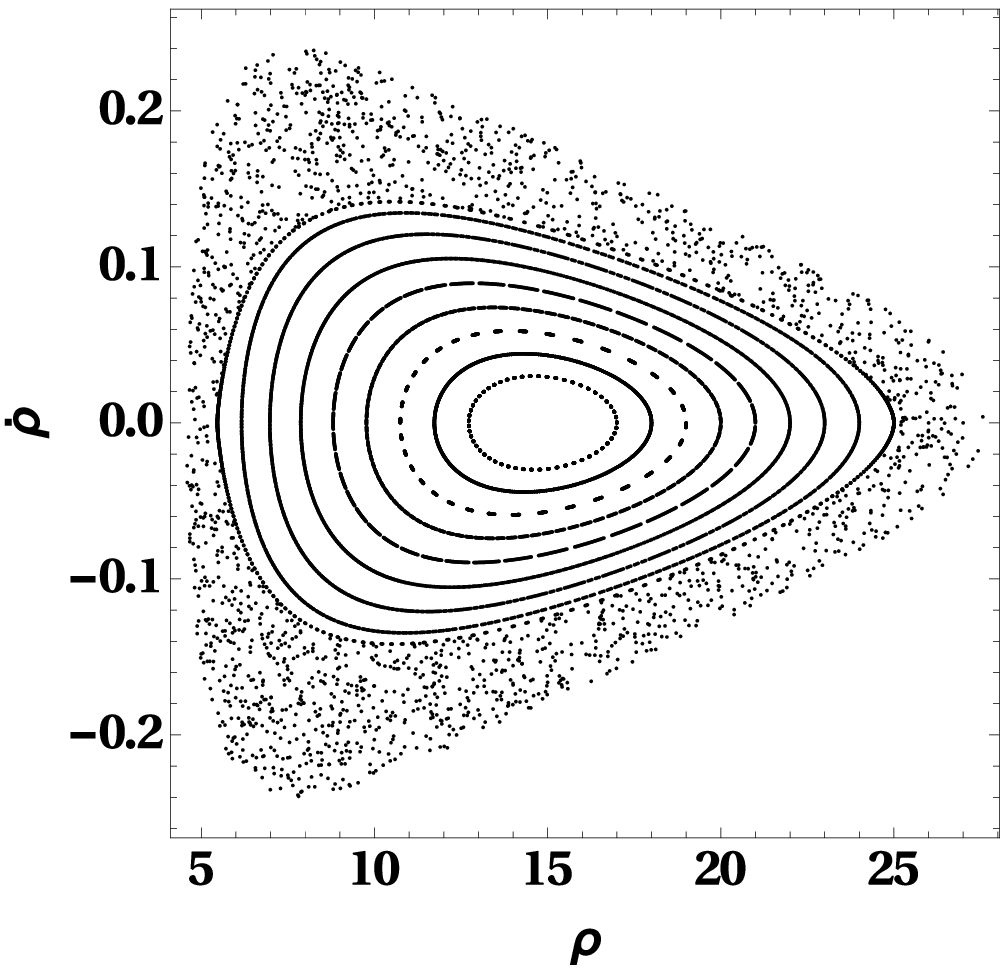}
   \includegraphics[width=0.45\textwidth]{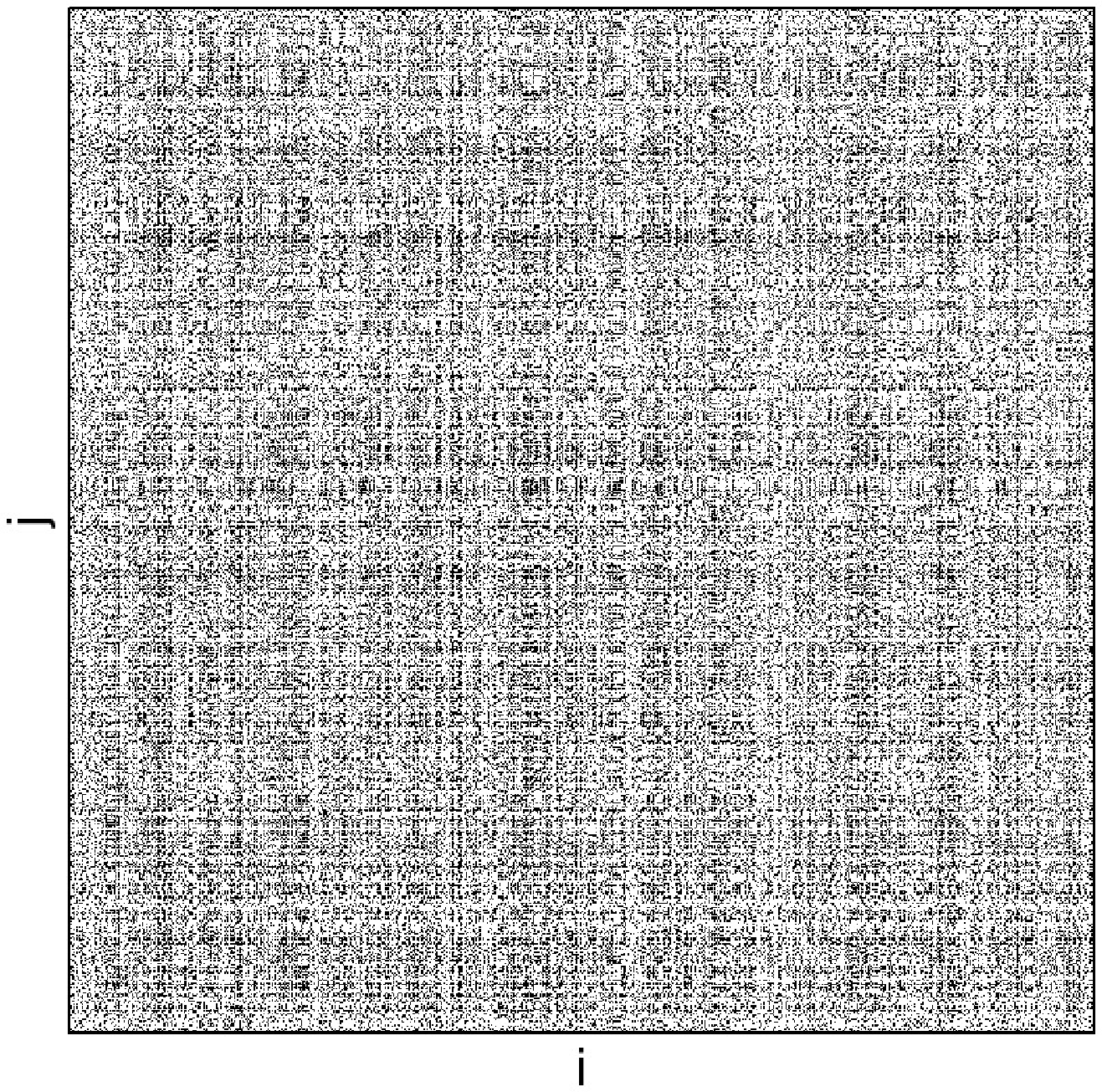}}
     \centerline{ \includegraphics[width=0.45\textwidth]{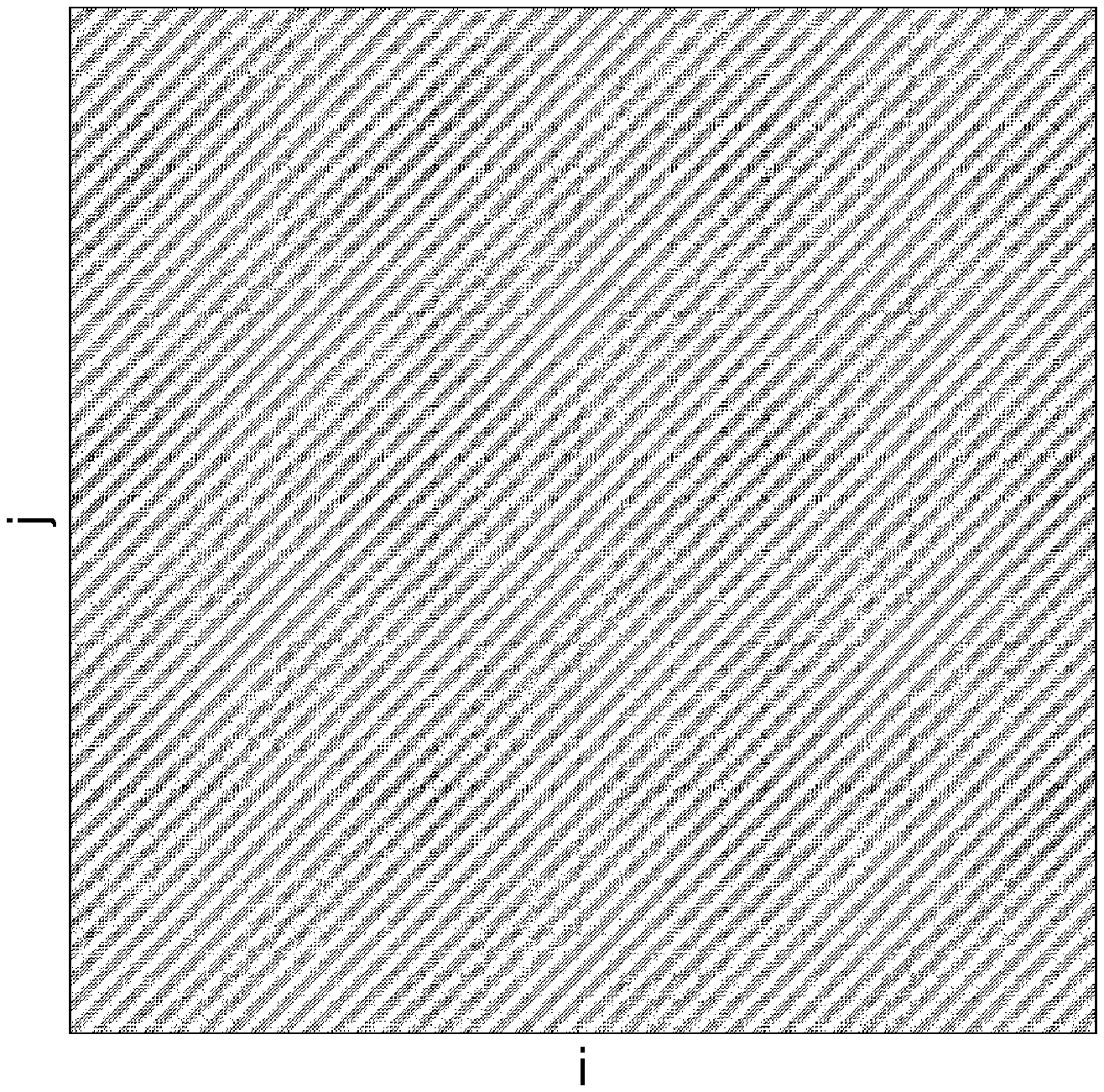}
   \includegraphics[width=0.45\textwidth]{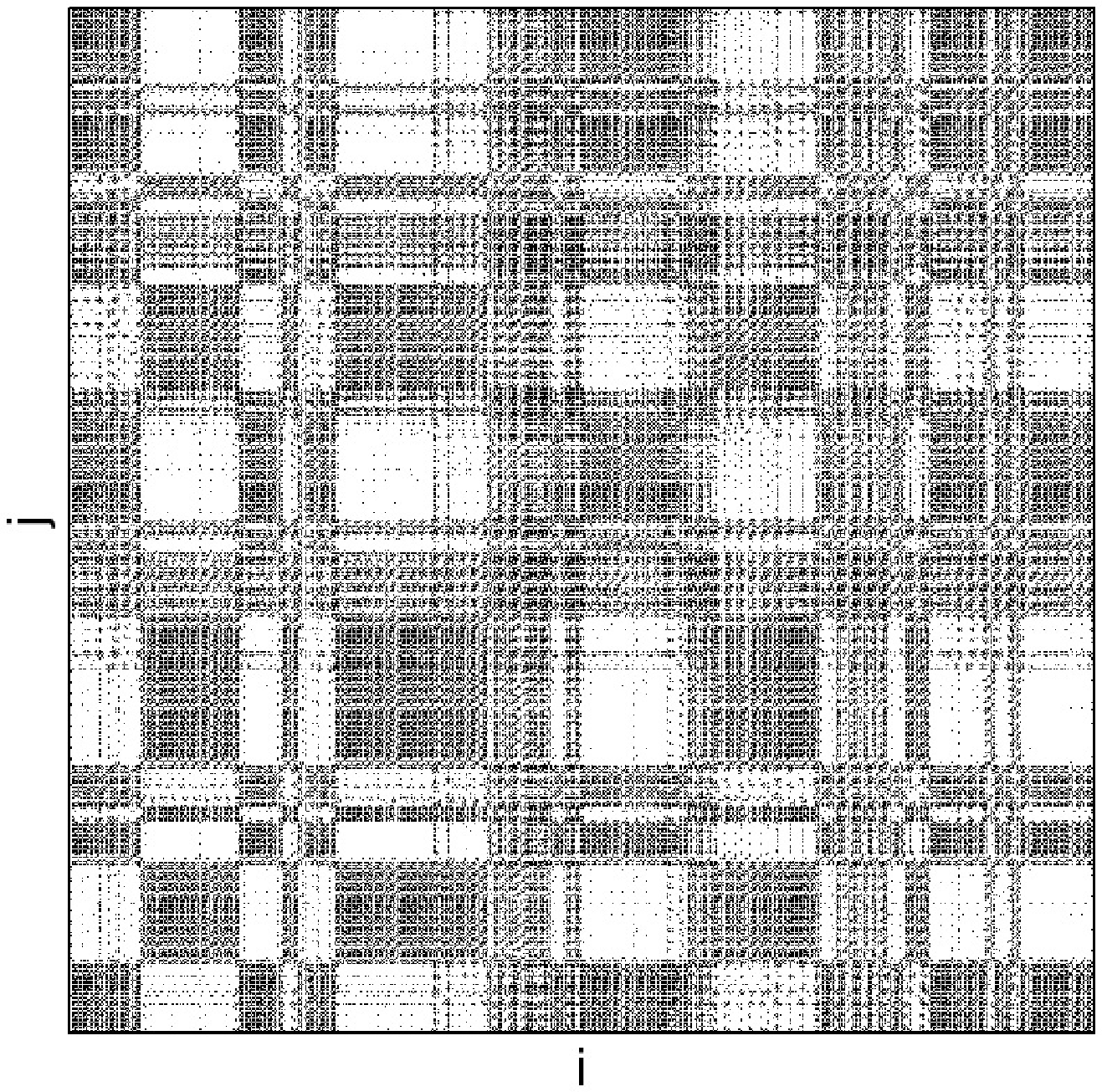}}
 \caption{The top left panel shows a Poincar\'{e} section recorded in the
 equatorial plane ($z=0$ and $\dot{z}>0$) for the set of trajectories with
 parameter values $E=0.97$, $L_z=3$, $a=0.6$, $M=1$, $b=3$. 
 The last three quantities define a spacetime for which 
 the quadrupole deviation parameter reads $q\approx 4.88>0$.
 This implies that the corresponding compact object is more prolate than a
 Kerr black hole.
 The top right panel shows the RP of the pure noise. 
 The bottom left is the RP of a regular orbit starting from
 $\rho=24,~\dot{\rho}=0$ on the Poincar\'{e} section. The bottom right is the
 RP of a chaotic orbit starting from $\rho=26,~\dot{\rho}=0$ on the Poincar\'{e}
 section.
  }
  \label{Fig:PoinSec}
 \end{figure}
 
 In order to show how deterministic motion can be discerned from noise by RP
 techniques we identify the different types of dynamics of geodesic orbits on a
 Poincar\'{e} section and provide the corresponding RP. In particular, the
 top left panel of  Fig.~\ref{Fig:PoinSec} shows a Poincar\'{e} section on the
 plane $(\rho,\dot{\rho})$ which includes the analyzed orbits. On this section 
 the chaotic sea (scattered points) surrounds KAM curves (homocentric closed
 curves) forming the main island of stability. According to the method described
 in Sec.~\ref{sec:LyapNum} the maximal Lyapunov number of the orbit belonging to
 the chaotic sea was found to be $\rm{mLCE}\simeq10^{-2.87}$. For numerical
 details of the calculation see Ref.~\refcite{geochaos}.
  
 Deterministic systems generally tend to produce characteristic linear patterns
 in a RP. Regular orbits typically form long diagonal lines, while
 deterministic chaos results in a more complex pattern, as shown in the bottom 
 left and bottom right panels of Fig.~\ref{Fig:PoinSec} respectively. Noise
 produces almost homogeneously distributed recurrence points (top right panel
 of Fig.~\ref{Fig:PoinSec}). Noise in this work is represented by the
 {\it Mathematica} function RandomReal$[{-x, x}]$, which gives a pseudorandom real
 number in the range $[{-x, x}]$, where $x$ is a real positive number.
 The $i,~j$~axes on the RP are representing the consecutive points of the time
 series. In other words $i,~j$ represent the time evolution, since $\tau$ is
 equal to $i~\Delta \tau$ and $j~\Delta \tau$, where $\Delta \tau$ is the time
 step at which the data points of the time series are registered.
 
 For individual cases a visual survey of RPs is a reliable method to discern 
 regular from chaotic motion, and determinism from noise. Nevertheless, in order
 to perform a systematic analysis of the measured or simulated data it is
 necessary to employ a more systematic approach based on the recurrence 
 quantification analysis (RQA). Indicators obtained by RQA provide various
 statistical measures of recurrences. Their definitions and basic properties can
 be found in the review Ref.~\refcite{marwan07}. In our analysis, we employ the indicator
 measuring the overall recurrence rate
 \begin{align} \label{eq:RR}
  \displaystyle RR(\varepsilon)=\frac{1}{N^2}\sum^{N}_{i,j=1} \mathbf{R}_{ij}(\varepsilon)\quad,
 \end{align}
 which determines the relative density of recurrence points in the RP constructed
 from the time series of $N$ data points, and the indicator
 \begin{align} \label{eq:DET}
 DET(\varepsilon)= \sum^{N}_{l=l_\textrm{min}} l P(\varepsilon,l) \biggm/
     \sum^{N}_{i,j=1} \mathbf{R}_{ij}(\varepsilon)\quad,
 \end{align}
 which reflects the amount of determinism present in the signal by measuring the
 ratio of the number of recurrence points forming diagonal lines in the RP
 (of the length at least $l_{\rm min}$) to the number of all recurrence points.
 Function $P(\varepsilon,l)$ represents the histogram of lengths $l$ of the diagonal lines in the RP:
 \begin{align} \label{eq:histog}
  P(\varepsilon,l)   =   \sum^{N}_{i,j=1}(1-\mathbf{R}_{i-1,j-1}(\varepsilon))
  (1+\mathbf{R}_{i+l,j+l}(\varepsilon)) \prod^{l-1}_{k=0}\mathbf{R}_{i+k,j+k}(\varepsilon) \quad. 
 \end{align}

 The above equations show that the numerical values of the RQA measures depend
 strongly on the value of recurrence threshold  $\varepsilon$, which has to be
 taken into account in the analysis. In general, $DET$ parameter is a reliable
 measure of determinism \footnote{Although in Ref.~\refcite{marwan11} it has been 
 shown that $DET$ may fail in some artificial non-physical systems.}. Recurrence analysis and RQA 
 have already been applied to detect the onset of chaos in various non-integrable
 relativistic systems \cite{kopacek10a, kopacek10b,SemSuk12,kovar13,SemSuk13,kopacek15}.
 In these systems the RQA indicator $DIV$ defined as an inverse value of the 
 length of the longest diagonal line found in the RP
 \begin{align} \label{eq:DIV}
  DIV=\frac{1}{\max_{i=1,...,N}\{l_i\}}
 \end{align}
 proved to be very useful due to its connection to Lyapunov exponents. In the
 present work, however, we combine the recurrence indicators $RR$ and $DET$ to obtain
 operational criterion allowing to detect the deterministic nature of (simulated)
 signals with noise. In Ref.~\refcite{sukova16} a similar criterion has been 
 recently applied to discern chaos from noise in signals from X-ray binaries.
 Here, this criterion is tested for the first time for the trajectories with the
 dissipation due to the gravitational radiation reaction (Sec.~\ref{sec:dissipation}).
 
 \section{RQA discerning determinism from noise}\label{sec:RQA}
 
 Noise affects the appearance of RPs and values of RQA indicators. Nevertheless,
 it has been shown that proper choice of the threshold parameter $\varepsilon$
 may minimize the impact of noise and recurrence analysis remains
 reliable for reasonable noise levels \cite{marwan07,thiel02}. 
 In the following we shall investigate what is the highest noise level which
 might be present in the signal so that its deterministic nature is still 
 detectable by means of recurrence analysis. 

 In our case, the noise is introduced by adding the function  RandomReal$[{-x, x}]$
 to each component of the orbit $\bm{y}=\{\rho,~\dot{\rho},~z,~\dot{z}\}$.
 By varying $x$ we set the level of noise we want to add to the signal. In order
 to study the degree of stochasticity in the signal, we define a noise-to-signal
 (NS) ratio as $NS\equiv x/\mu({|y_i|})$ by taking the mean value for each  component
 of the analyzed temporal segment of the trajectory. In fact, noise-to-signal $NS$
 is just the inverse value of the commonly used signal-to-noise ratio $SN$, i.e.
 $NS\equiv SN^{-1}$, but the former is more appropriate in our context, since
 we investigate the effect of gradually increasing noise level in simulated data.
  
 Recurrence analysis is a robust method highly suitable for experimental data as
 it accepts the data in a raw state and still detects fundamental properties of
 underlying dynamical system which are non-trivially encoded in the signal.
 In the case of a gravitational wave signal from an EMRI we expect to obtain a 
 one dimensional interferometric data series for the analysis. This series will
 represent the spacetime perturbation as detected by the apparatus. However,
 the characteristic frequencies of the trajectory of the inspiraling object are 
 encoded in the spacetime perturbation modes $h_{+}$ and $h_{\times}$. For 
 example, the correlation between the spacetime perturbation modes and the
 components of the trajectory is implied in Ref.~\refcite{Kiuchi04} and explicitly shown
 in Figs. 8 and 9 of Ref.~\refcite{Gair08} for $h_{+}$. Having this in mind, one can 
 directly investigate the trajectory of the inspiral instead of the waveforms of
 the perturbation modes. Since the recurrent patterns of the
 trajectory inherently reflect frequencies of the underlying dynamical system,
 it is reasonable to assume that the conclusions found in Ref.~\refcite{Gair08} 
 for standard frequency analysis tools remain valid also for the recurrence analysis,
 and thus, we can use the trajectory directly instead of the simulated waveforms
 derived from this trajectory.
 
 We performed the analysis for each component of the trajectory 
 separately, i.e. we treated each component as an independent one dimensional time
 series. Nevertheless, we also checked the case when the components are used
 simultaneously. We found that the analysis with such a varied data sets led to 
 equivalent, and similarly indicative, results. This outcome was expected since
 the MSM spacetime does not provide separable system (unlike Kerr spacetime).
 Therefore, without loss of generality, we discuss here the results based on
 the $\rho$ component only. Note that the analysis was performed in a
 reconstructed phase space as discussed in appendix~\ref{sec:RQAparam}.

 \begin{figure} 
  \centerline{ 
   \includegraphics[width=0.25\textwidth]{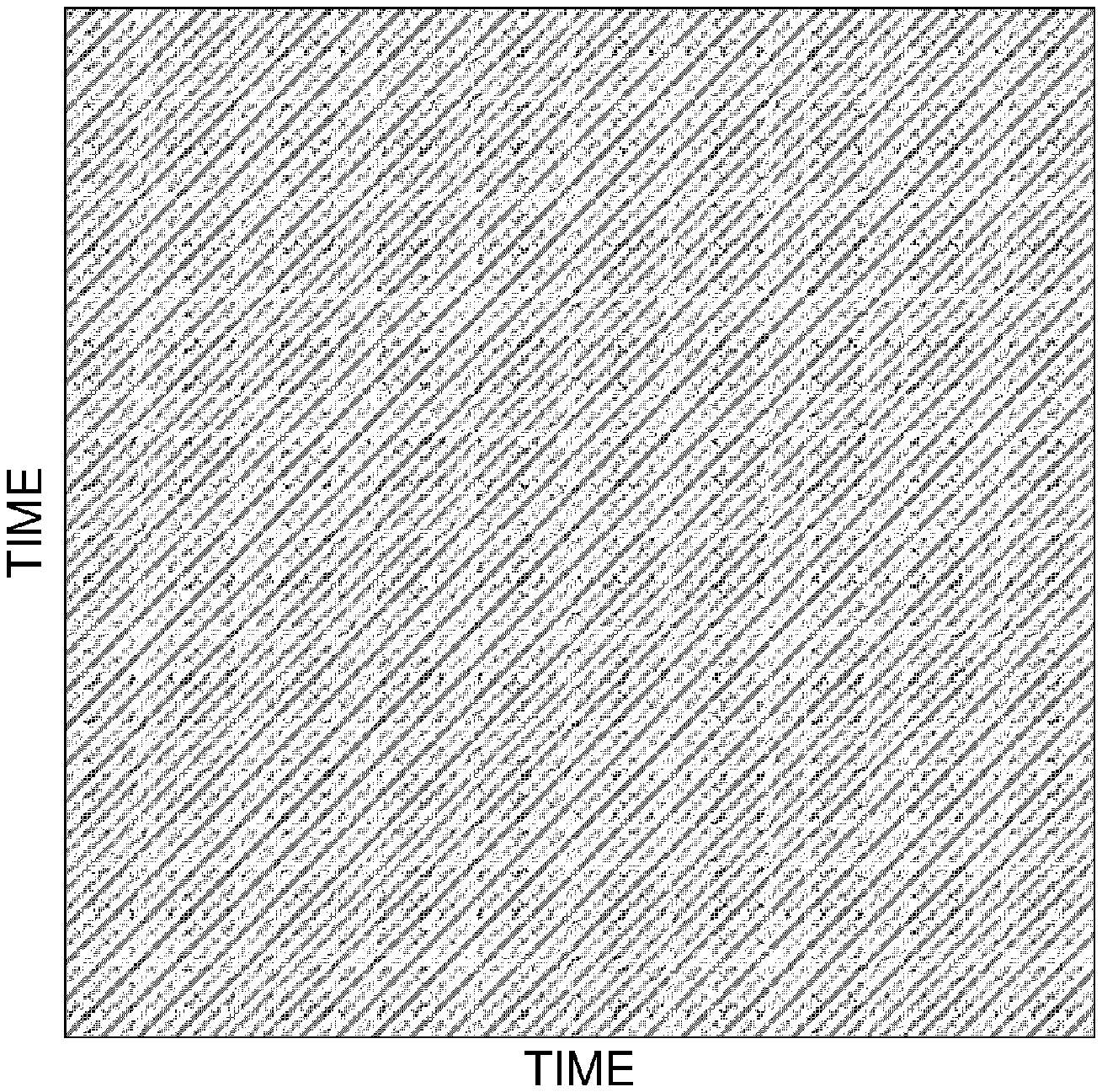}
   \includegraphics[width=0.25\textwidth]{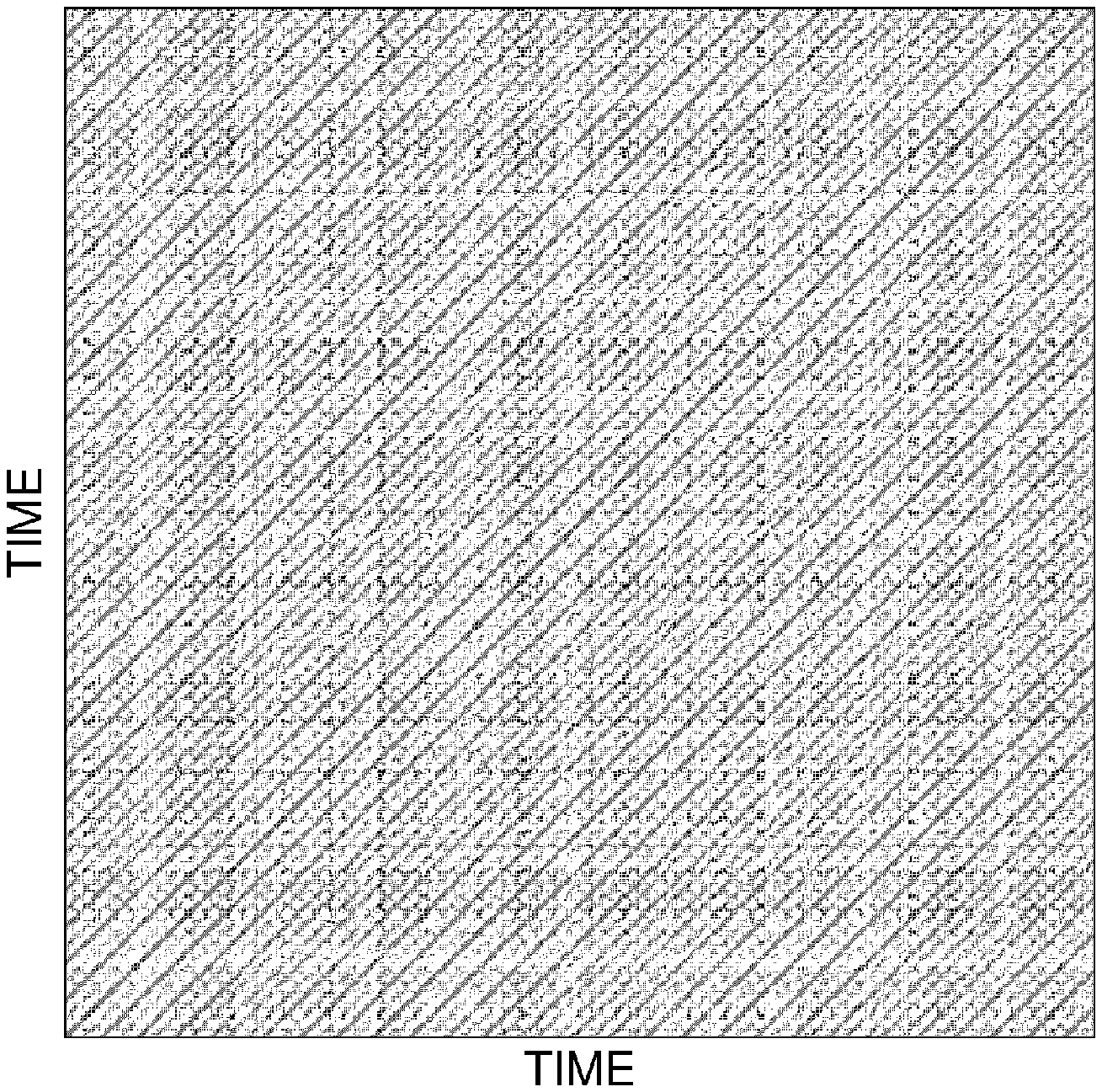}
   \includegraphics[width=0.25\textwidth]{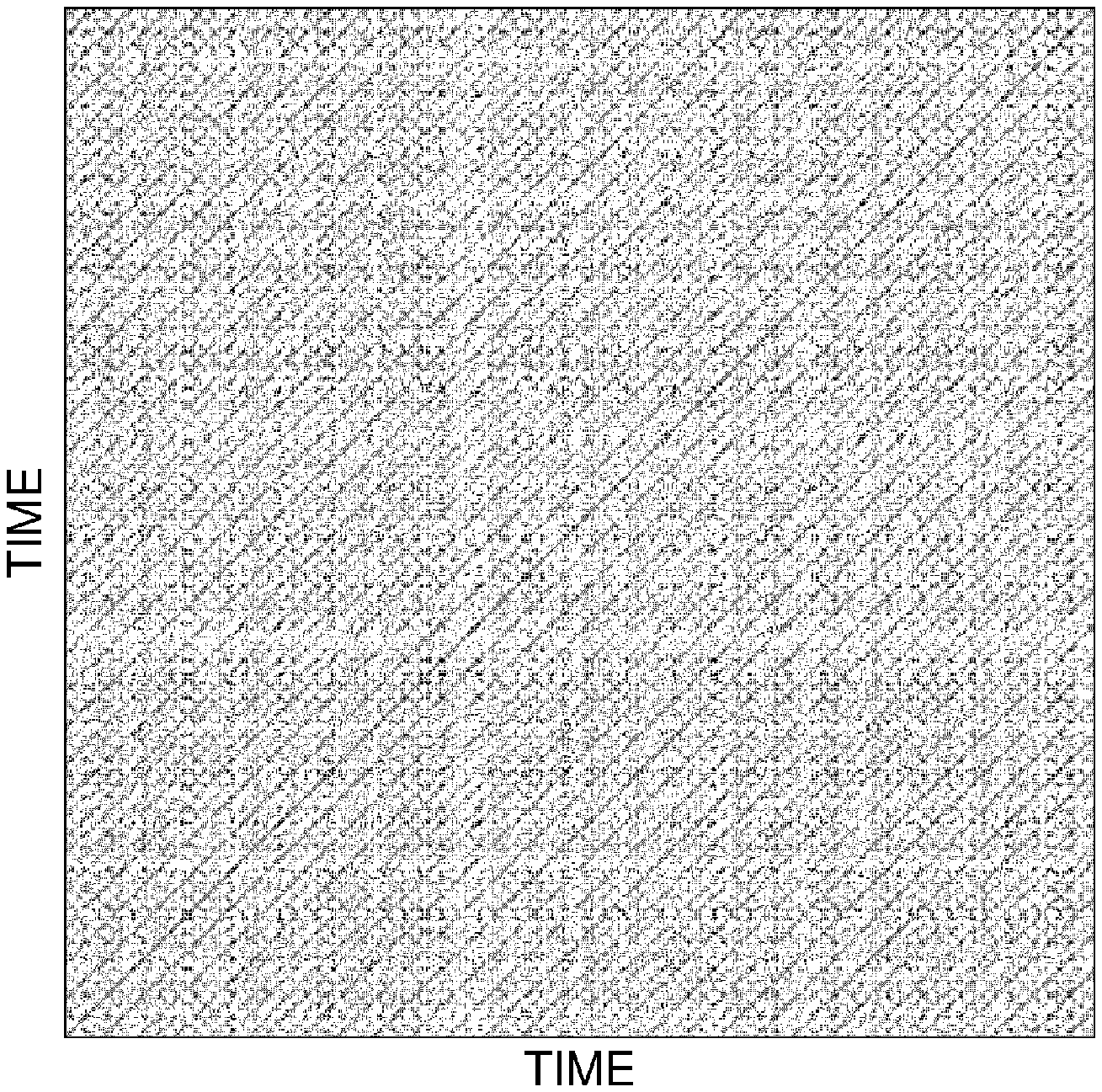}
   \includegraphics[width=0.25\textwidth]{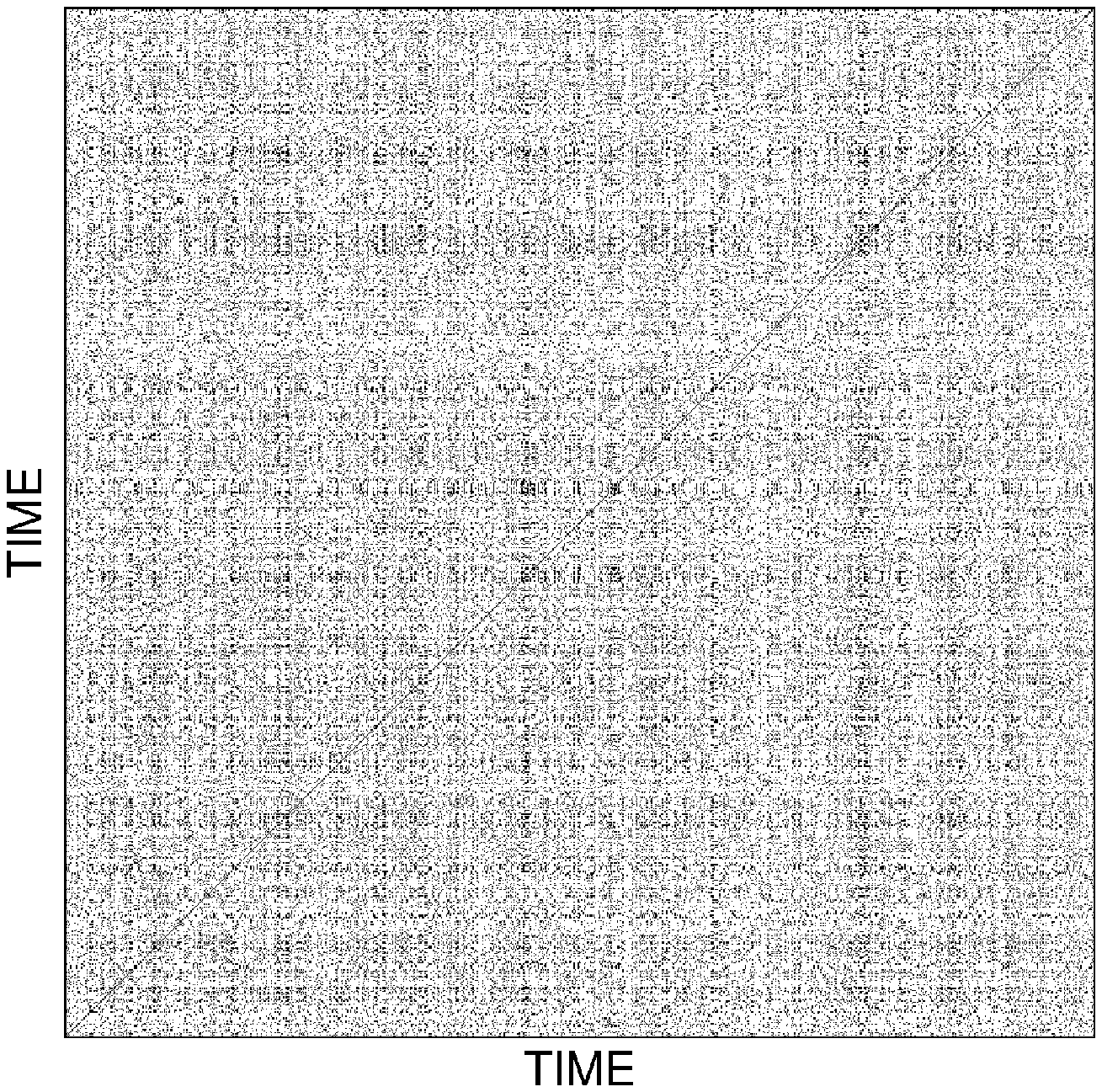}
   }
   \caption{A series of RPs for threshold $\varepsilon=1.8$ corresponding to the regular
   trajectory of Fig.~1 with the amount of noise increasing from left to
   right. The first panel has zero noise, the second has $NS\approx10\,\%$, the third has
   $NS\approx15\,\%$, and the last panel shows pure noise. We see characteristic diagonal
   pattern typical for regular trajectories being gradually buried in the increasing noise.}
  \label{RP_nodissip_reg}
 \end{figure}
 
  \begin{figure} 
  \centerline{ 
   \includegraphics[width=0.25\textwidth]{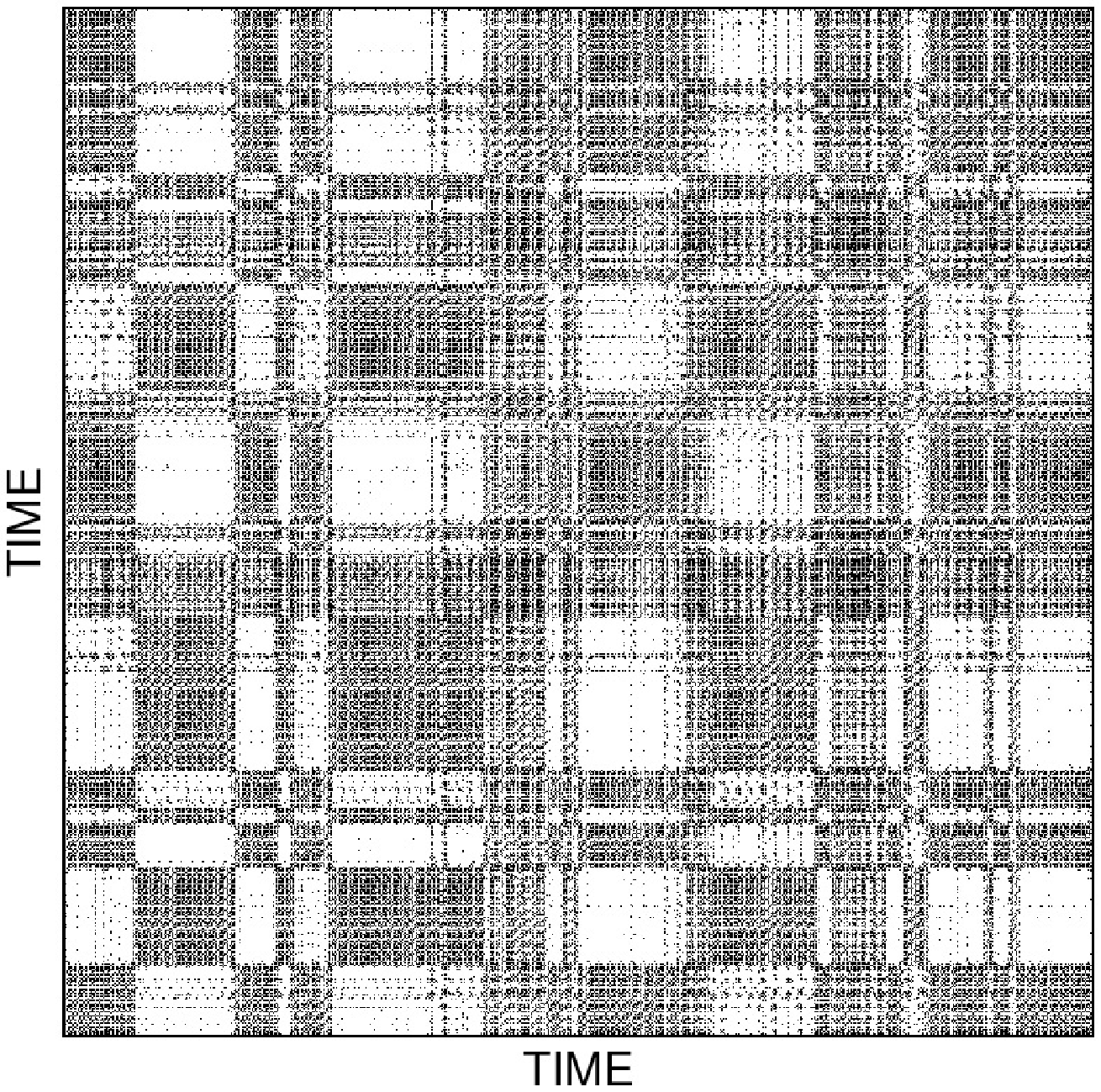}
   \includegraphics[width=0.25\textwidth]{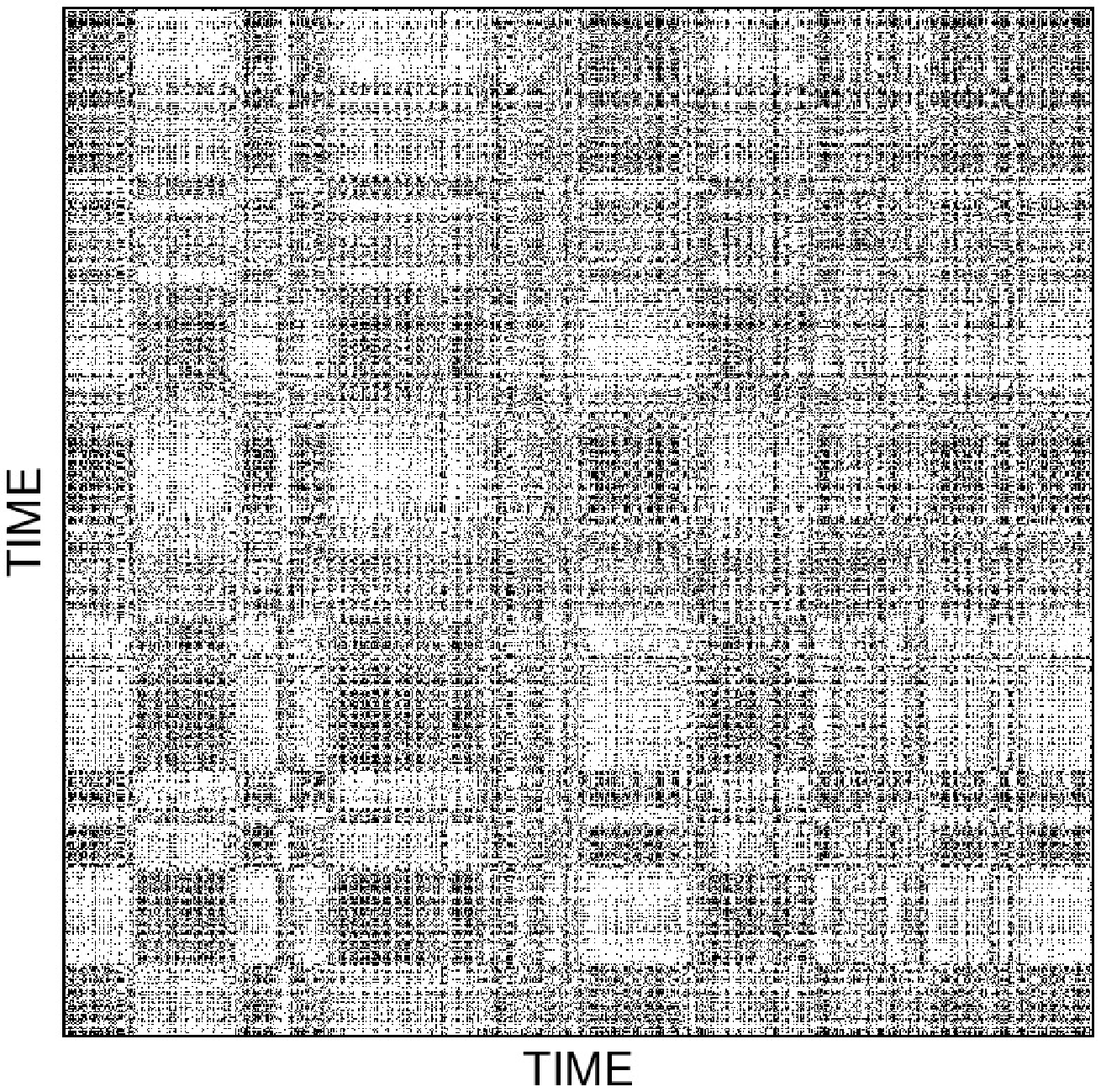}
   \includegraphics[width=0.25\textwidth]{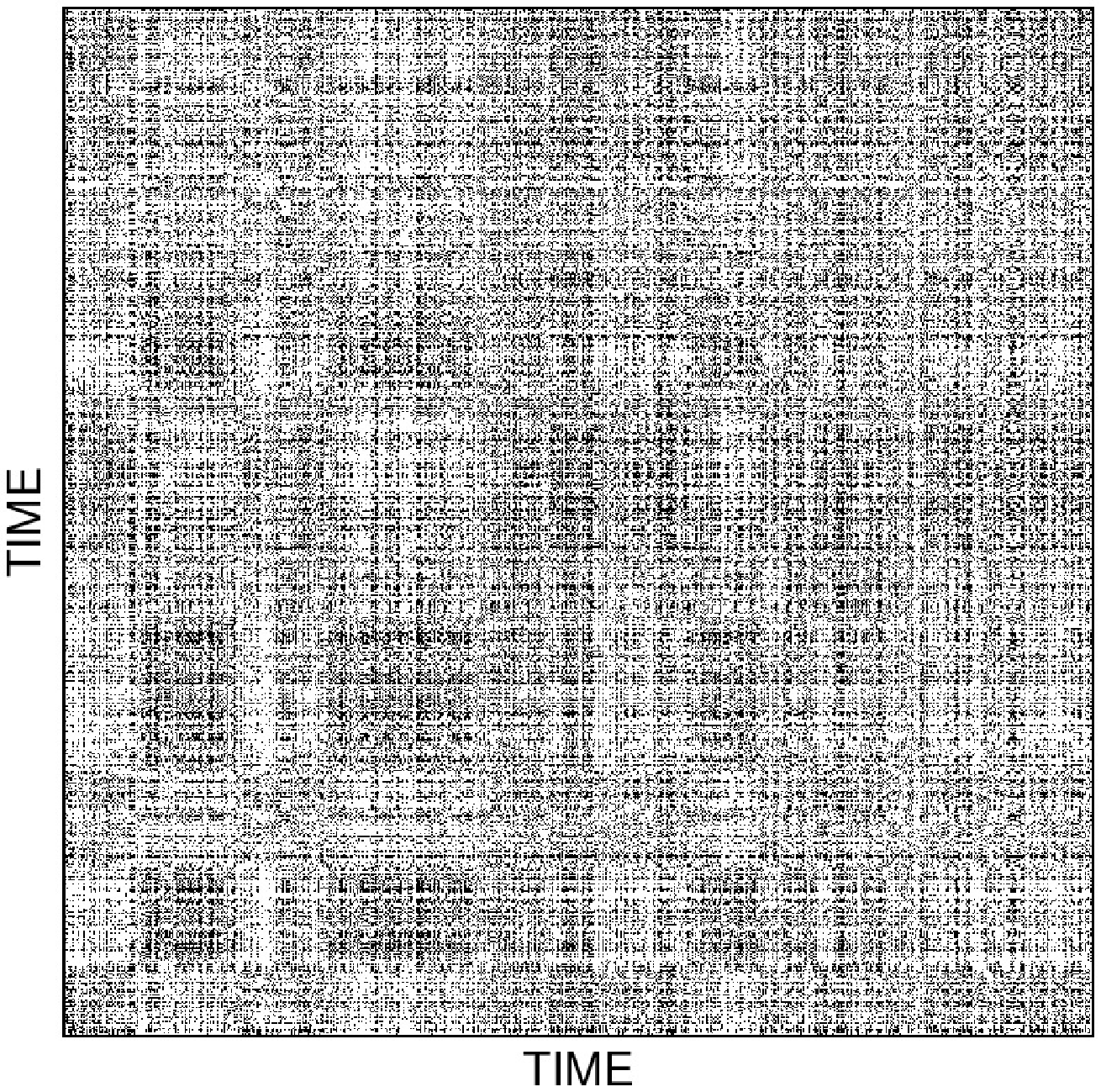}
   \includegraphics[width=0.25\textwidth]{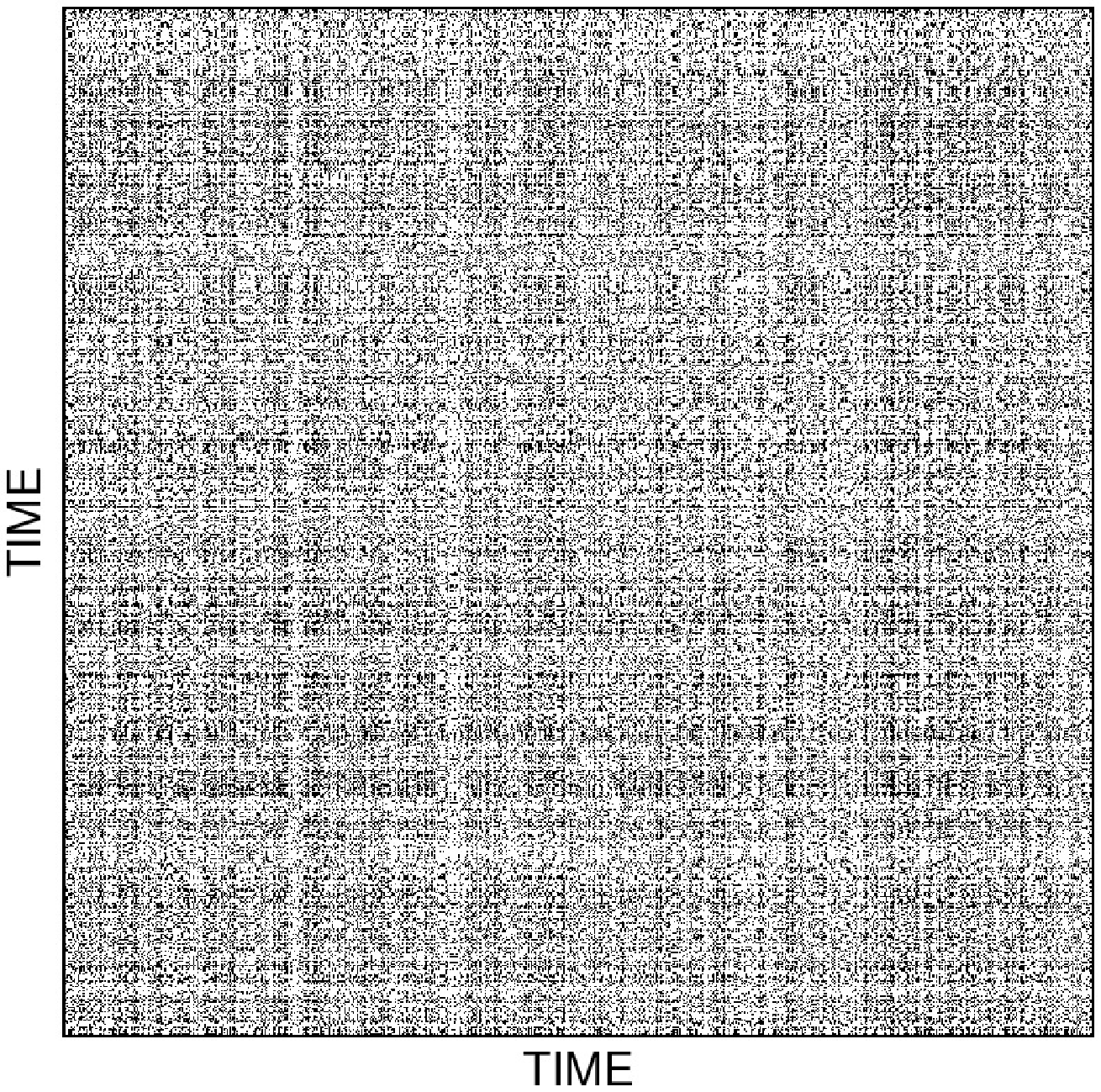}
   }
   \caption{A series of RPs for threshold $\varepsilon=2.15$ corresponding to 
   the chaotic trajectory of Fig.~1 with the amount of noise increasing
   from left to right. The first panel has zero noise, the second has
   $NS\approx5\,\%$, the third has $NS\approx40\,\%$, and the last panel
   shows pure noise. We see the recurrence patterns typical for chaotic trajectories
   being gradually buried in the increasing noise.}
  \label{RP_nodissip}
 \end{figure}

  \begin{figure} 
  \centerline{ 
    \includegraphics[width=0.5\textwidth]{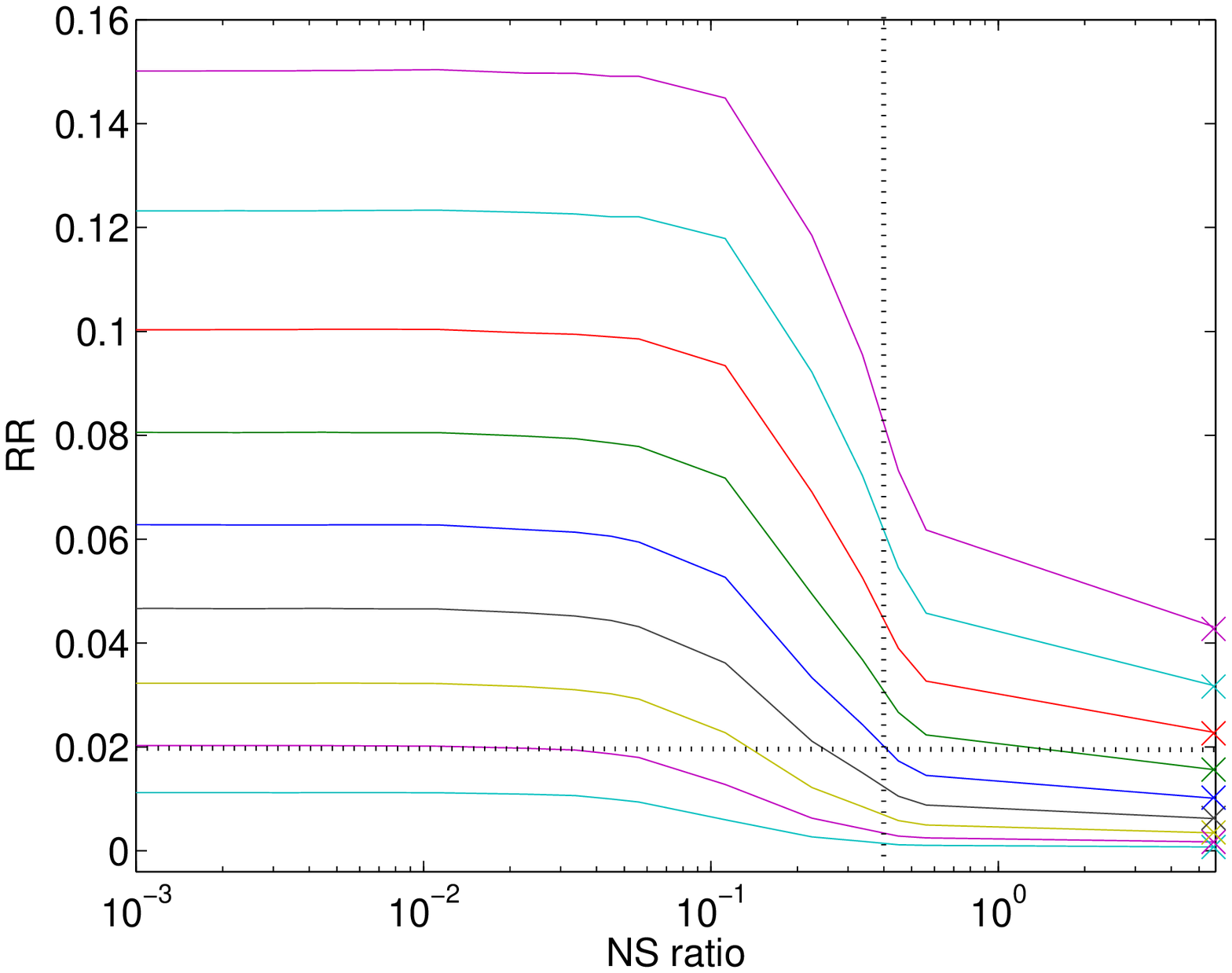}
     }
      \centerline{     
    \includegraphics[width=0.5\textwidth]{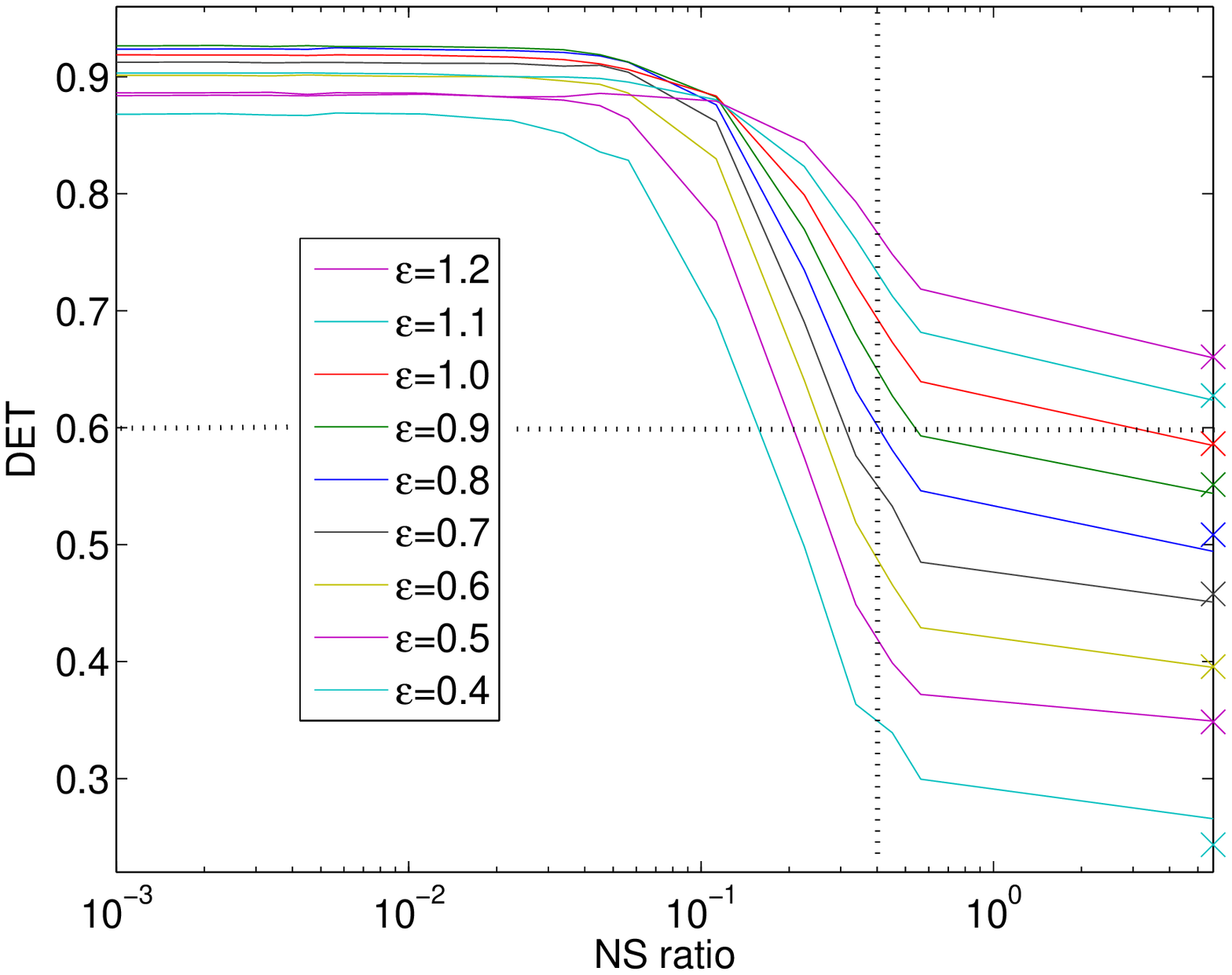}
  }
  \caption{The recurrence indicators $RR$ and $DET$ as a function of NS ratio for
  the chaotic trajectory. Each indicator is evaluated for several values of 
  $\varepsilon$ denoted in the legend of the $DET$-plot and corresponding also to
  the $RR$-plot. The curves for different values of $\varepsilon$ create an
  ordered sequence in the both plots -- lower $\varepsilon$ gives lower 
  recurrence rate and also lower value of $DET$.  The asymptotic values of the
  indicators evaluated for the pure noise are marked on the right vertical
  axes by cross-marks in the corresponding colors. The critical noise level of
  $NS=40 \%$ is denoted by the vertical dashed line.}
  \label{RQA_nodissip}
 \end{figure}

 \subsection{Geodesic case} \label{sec:geodesic}
 
 We begin with the analysis of orbits in the geodesic scenario in which the
 dissipation is neglected. In paricular, in Fig.~\ref{RP_nodissip_reg} we inspect
 RPs of regular trajectory (the one introduced in Fig.~\ref{Fig:PoinSec}) with
 gradually increasing noise. We observe that diagonal patterns characteristic for
 regular motion are visible
 for the noise levels up to $NS\approx15\,\%$. For higher noise levels the 
 visual detection of the deterministic component of the signal becomes ambiguous.
 In Fig.~\ref{RP_nodissip} we compare RPs of the chaotic geodesic trajectory
 (shown in the bottom right panel of Fig.~\ref{Fig:PoinSec}) with increasing
 noise level. We identify recurrence patterns typical for deterministic chaos
 which are gradually dissolving in the noise. We may safely identify these
 deterministic patterns
 for the values of noise up to $NS\approx40\,\%$. 
 Thus, the visual analysis of RPs allows to detect the deterministic nature of
 the chaotic system, although the signal is considerably contaminated by the noise.
 While in some other respects the chaotic dynamics may mimic behavior of stochastic
 systems, here we observe the opposite: by means of visual survey of RPs the
 deterministic ingredient of the signal may be distinguished more easily from
 the stochastic noise if it is generated by chaotic (rather than regular) dynamics.
 
 Nevertheless, for a systematic analysis we need a reliable quantitative
 criterion to detect determinism obscured by stochastic noise. This can be
 achieved by evaluating the dependence of $RR$ and $DET$ on the noise level for a
 sequence of values of recurrence threshold $\varepsilon$. In
 Fig.~\ref{RQA_nodissip} we present the results for the chaotic trajectory.
 The values of $RR$ and $DET$ generally decrease with decreasing
 $\varepsilon$. On the other hand, the values of both the indicators also
 decrease with increasing noise level which is not known {\it a priori}.
 Therefore, we first need to conveniently fix $\varepsilon$ (which is a free
 parameter in the analysis) so that we can use $DET$ to efficiently detect
 determinism in the signal with generally unknown amount of noise. 
 
 We suggest the following {\it operational criterion}: if the value of $DET$ is
 greater or equal to $0.6$ for such a value of $\varepsilon$ which gives
 $RR=0.02$, then the signal contains deterministic component. These threshold
 values have been selected {\it ad hoc} in such a way that on the one hand we 
 safely avoid false identification of the pure noise as deterministic, but on 
 the other hand we maximize the noise level for which we still correctly identify
 the deterministic component of the signal. For the general time series containing
 unknown amount of noise the algorithm we propose is to tune $\varepsilon$ to 
 obtain $RR=0.02$ and check the value of $DET$ evaluated with this
 $\varepsilon$. Applying this criterion in Fig.~\ref{RQA_nodissip} we see that 
 determinism in the analyzed data may be identified for the values of NS at
 least up to $\approx 40\,\%$ before the corresponding $DET$-curves 
 (i.e., those with $\varepsilon$ which gives $RR=0.02$ at given noise level) 
 start to drop below $DET=0.6$. In particular, in Fig.~\ref{RQA_nodissip} we
 observe that at noise level $NS\approx40\,\%$ the condition $RR=0.02$ is met
 by the blue curve evaluated with $\varepsilon=0.8$ and at the same noise level
 the corresponding blue $DET$-curve just meets the boundary value of $DET=0.6$.
 With more noise the relevant $\varepsilon$ would lead to $DET<0.6$.

   \begin{figure} 
  \centerline{ 
   \includegraphics[width=0.25\textwidth]{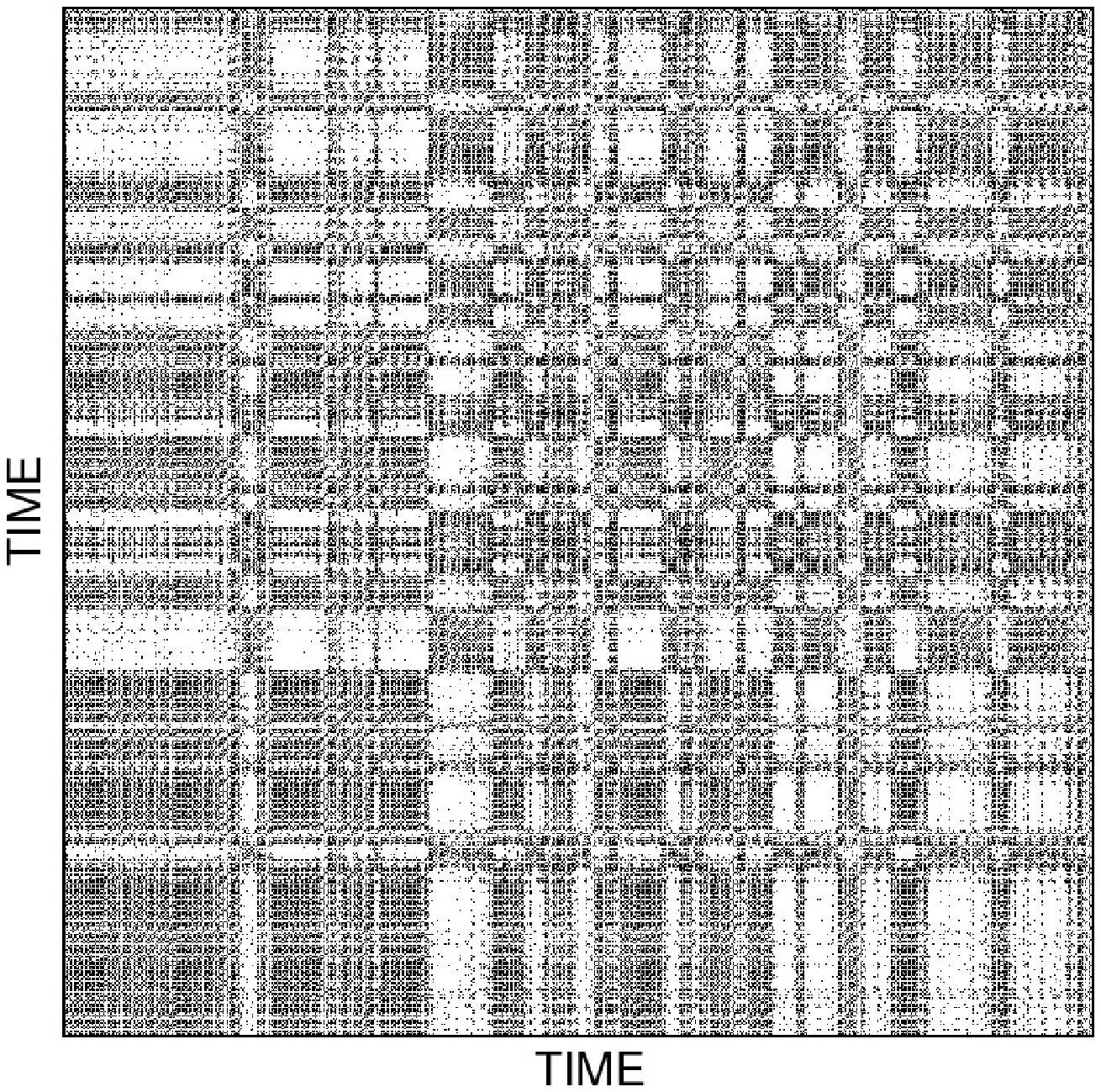}
   \includegraphics[width=0.25\textwidth]{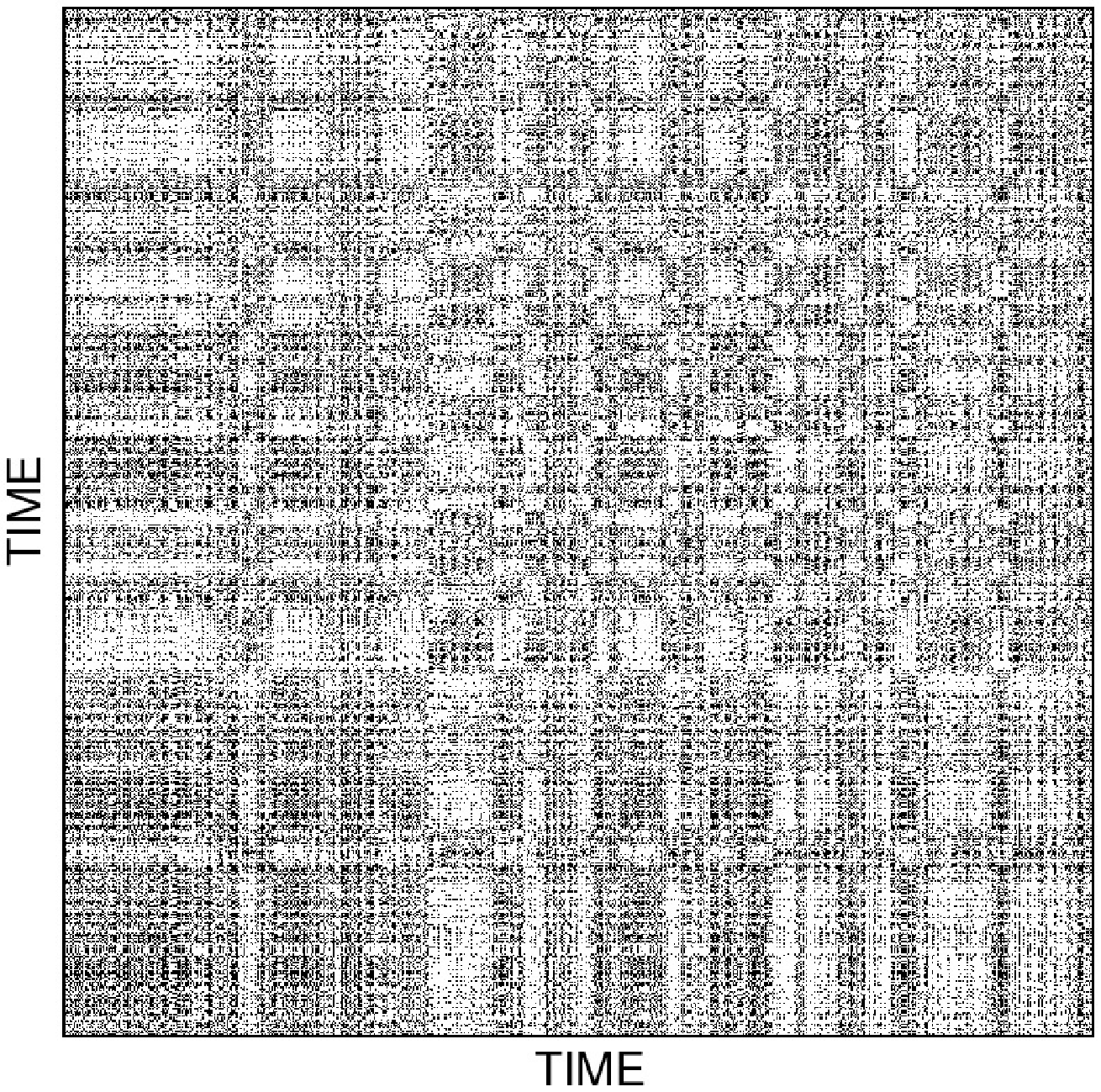}
   \includegraphics[width=0.25\textwidth]{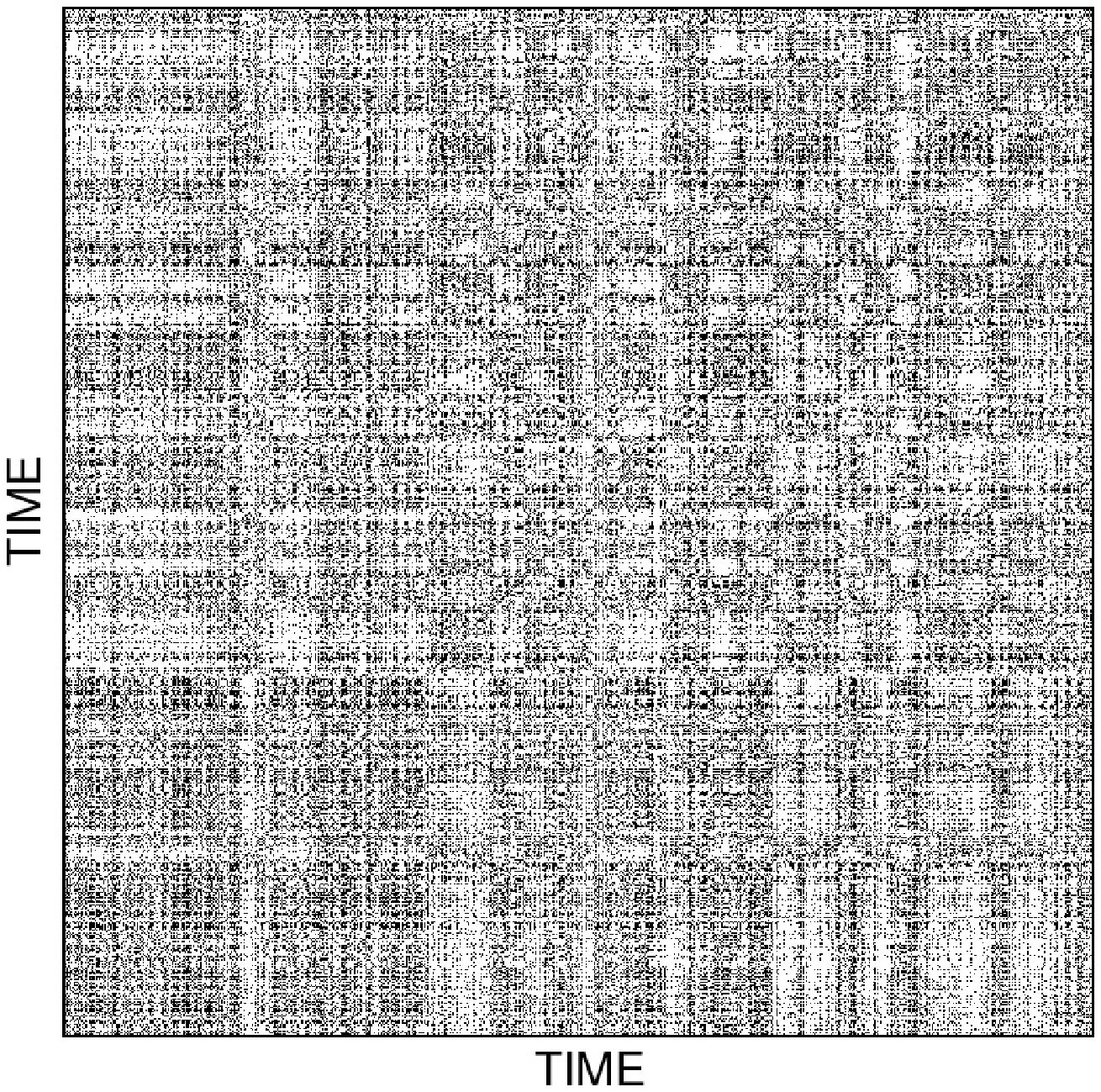}
   \includegraphics[width=0.25\textwidth]{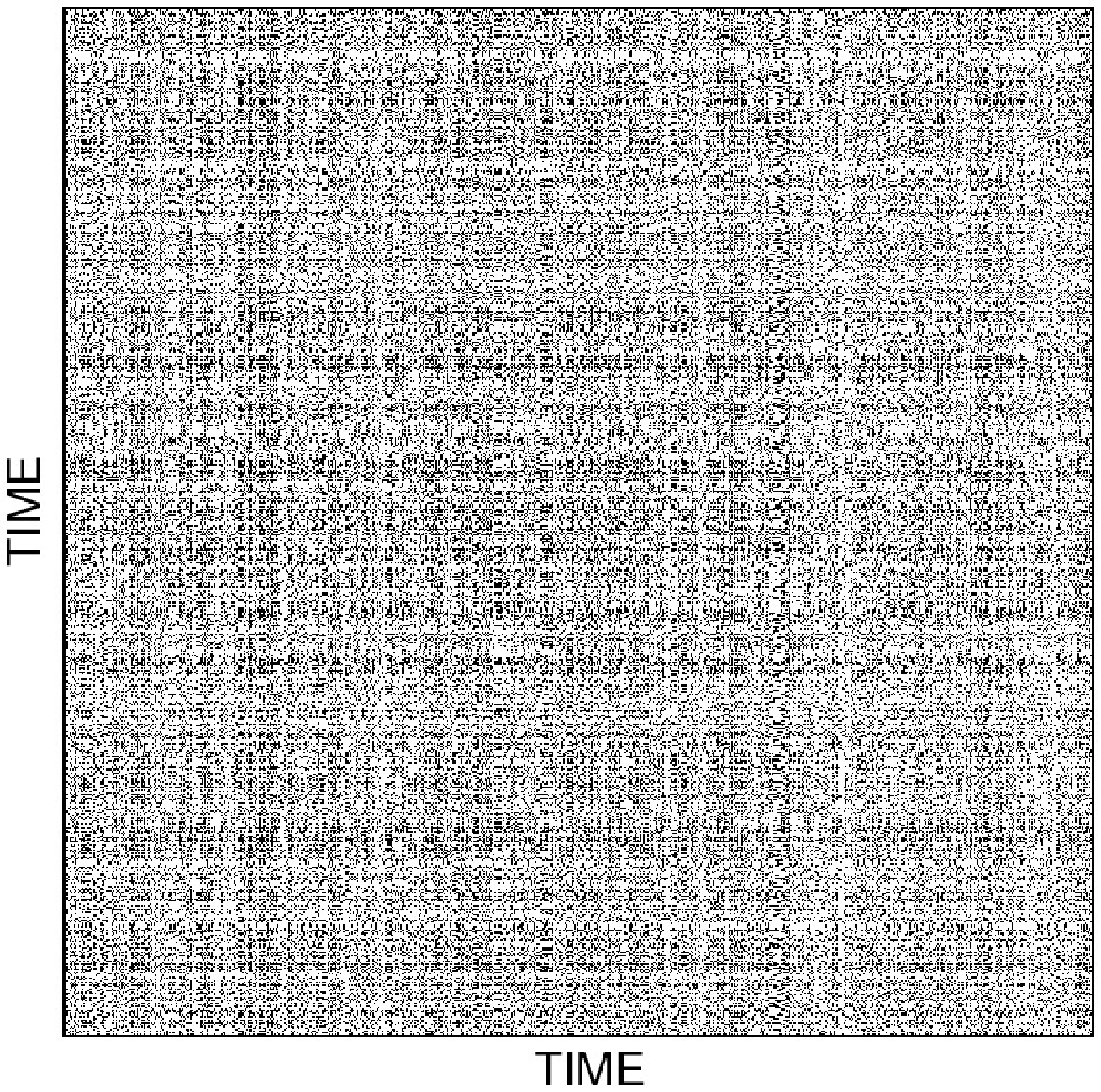}
  }
   \caption{A series of RPs for threshold $\varepsilon=2.15$ corresponding to a
   dissipating chaotic trajectory ($\nu=10^{-4}$) with increasing amount of noise
   from left to right. The first panel has zero noise, the second has $NS\approx3\,\%$,
   the third has $NS\approx20\,\%$, and the last panel shows pure noise. We see 
   the chaotic patterns being gradually buried in the increasing noise.
   }
  \label{RP_dissip}
 \end{figure}
 
  \begin{figure}
  \centerline{ 
   \includegraphics[width=0.5\textwidth]{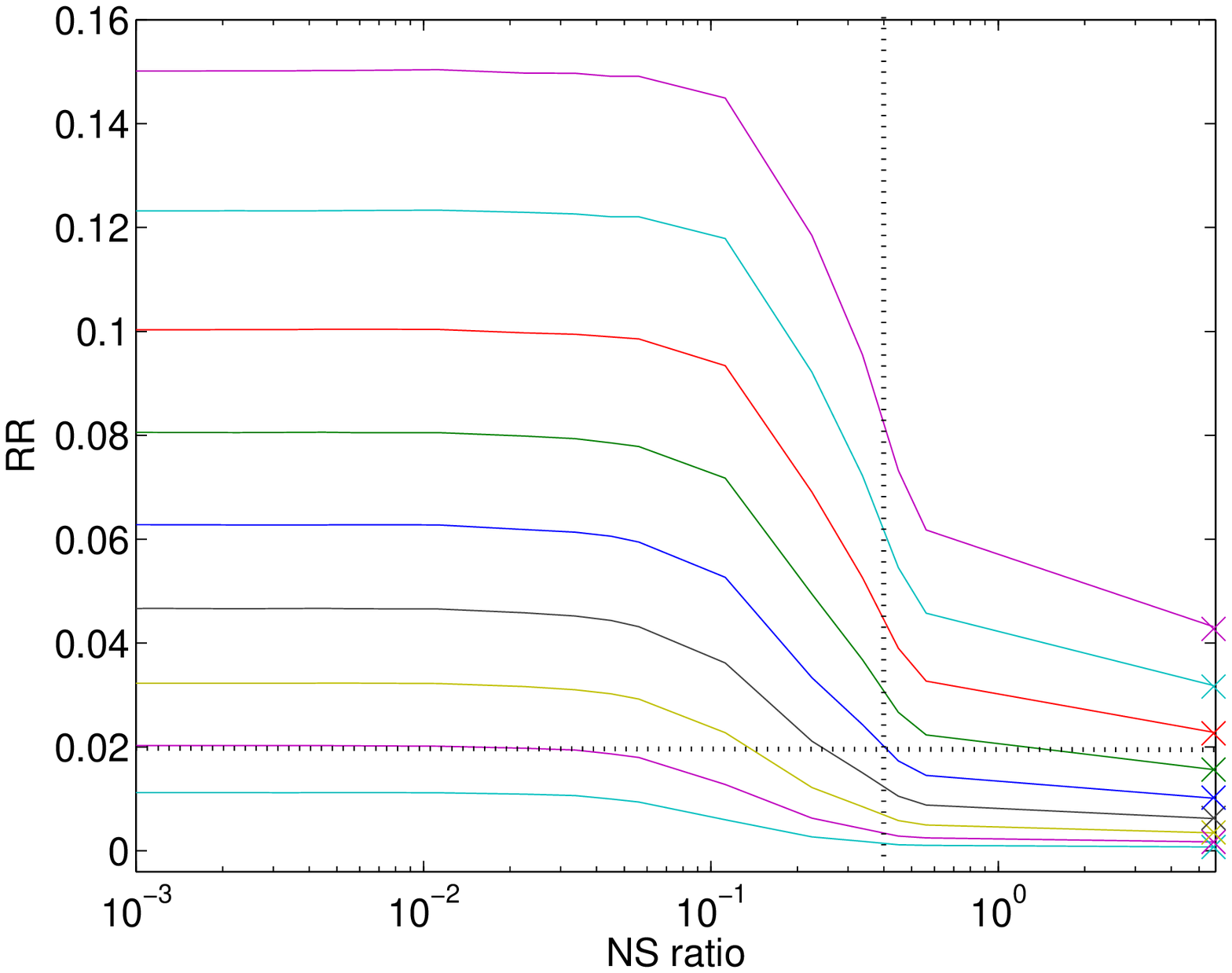}
   }
    \centerline{ 
   \includegraphics[width=0.5\textwidth]{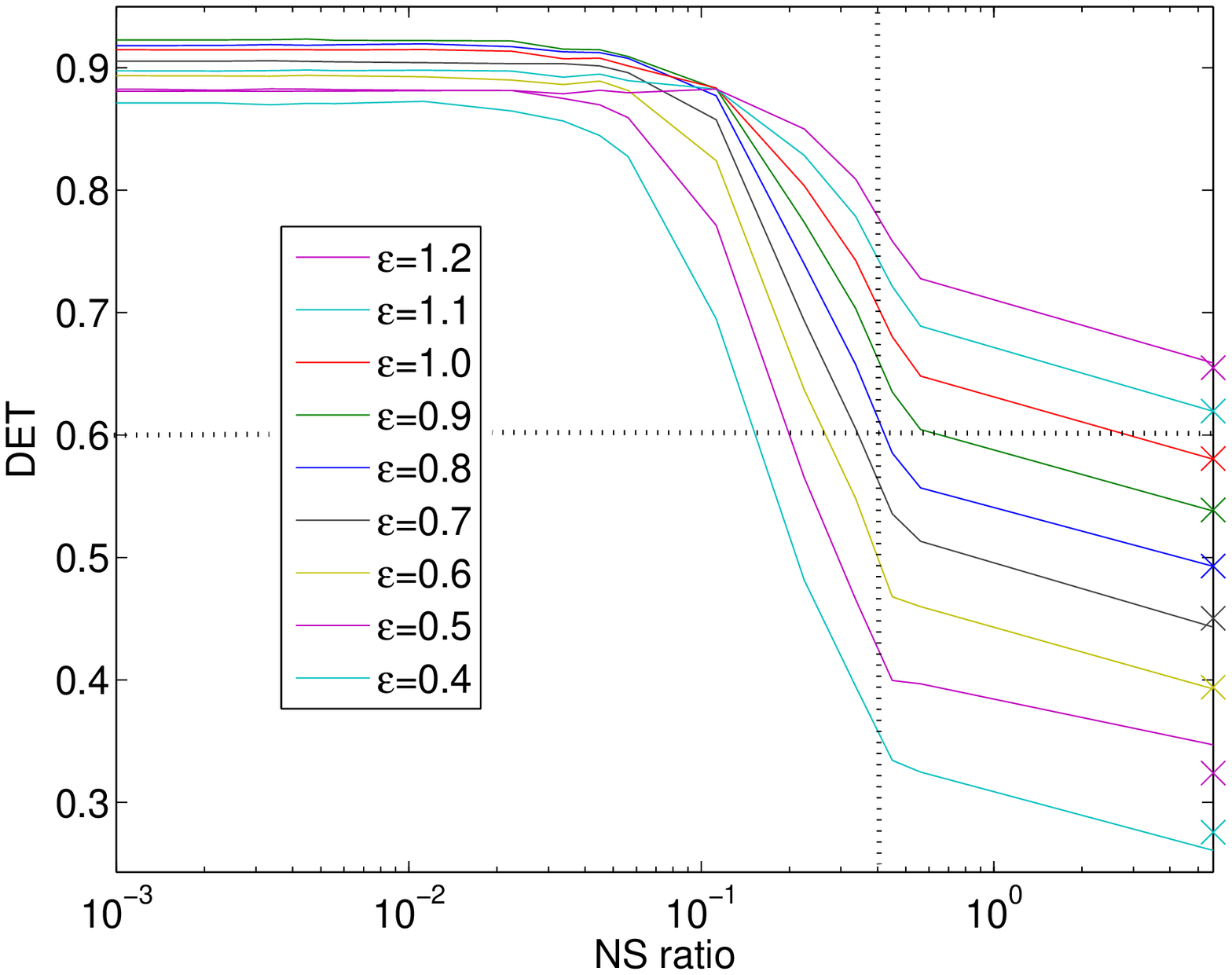}
  }
  \caption{The recurrence indicators $RR$ and $DET$ as a function of NS ratio for a
  dissipating chaotic trajectory. Each indicator is evaluated for several values of
  $\varepsilon$ denoted in the legend of the $DET$-plot. The critical noise level
  of $NS=40 \%$ is denoted by the vertical dashed line.}
  \label{RQA_dissip}
 \end{figure}

 \subsection{Inspiraling case} \label{sec:dissipation}
 
 To make our investigation more realistic, we have to include the adiabatic
 dissipation due to the radiation reaction into our evolution scheme. To achieve
 this we follow the approximative approach employed in Refs.~\refcite{nonKerr,Gair08}.
 In this scheme, the mass ratio $\nu=m/M$ of the EMRI adjusts the applied
 dissipation rate, where m is the mass of the inspiraling body.
 The higher $\nu$ is, the higher the applied dissipation will
 be. In general, as the dissipation increases an inspiral shifts faster between
 geodesic orbits. Thus, by increasing the dissipation the recurrence rate is
 reduced, which implies that the recurrence analysis has a certain limit of 
 applicability. The question is whether this limit is sufficient for an EMRI or
 not.
 
 The upper limit of the mass ratio for an EMRI is $\nu=10^{-4}$ \cite{LISA}. Thus,
 by checking this upper limit, we apply the highest rate of dissipation on the 
 inspiral. Moreover, we intentionally choose the chaotic case which maximally 
 differs from the scenarios covered by currently existing templates, since the
 current templates are based on the assumption of an integrable dynamical
 background which does not allow chaotic dynamics. By combining chaotic motion
 in an EMRI and the dissipation with high value of $\nu$ we obtain the worst case
 scenario for the current templates and we test whether the recurrence method
 could be used under these circumstances. In particular, in Fig.~\ref{RP_dissip}
 we compare RPs of chaotic dissipative trajectory with increasing amount of noise.
 The typical patterns of deterministic chaos are still partially visible even if
 the noise increases to $NS\approx20\%$. Thus, the deterministic nature of the
 signal with high noise level may still be visually identified in RPs, even if 
 the upper limit of mass ratio is considered. For the regular orbit with
 dissipation, the limit is $NS\approx15\%$ (we do not present the plots for this
 case). Thus, as in the geodesic case, the visual detection of determinism 
 appears to be more effective for chaotic trajectories.
 
 Applying our criterion on values of RQA indicators $RR$ and $DET$, we find that the
 deterministic nature of the signal may be detected for the noise levels up to
 $NS\approx40\,\%$ for the chaotic trajectory with dissipation 
 (see Fig.~\ref{RQA_dissip}). Although the appearance of RPs and visual
 stability of their characteristic patterns is considerably affected by the
 dissipation (compare Figs.~\ref{RP_nodissip} and \ref{RP_dissip}), the behavior
 of $RR$ and $DET$ indicators is almost indifferent to the dissipation (compare 
 Figs.~\ref{RQA_nodissip} and \ref{RQA_dissip}) and the critical noise level
 $NS\approx40\,\%$ is comparable to the geodesic case discussed in
 Sec.~\ref{sec:geodesic}. Moreover, unlike in the case of visual detection of
 determinism using RPs, here the chaotic orbits are not privileged, i.e.
 the noise to signal limit is the same for chaotic and regular orbits.
 
 Moreover, we believe that the values of the critical noise levels
 mentioned above are in fact the lower limits. The theoretical reliability of
 the method is probably considerably higher and its performance could be
 enhanced by further tuning. 
 
 \section{Conclusions}\label{sec:Conc}
 
 We have shown that visual survey of recurrence plots allows to detect
 determinism in the signal with considerable amount of noise. We have formulated
 a quantitative criterion of determinism of the signal based on the combination
 of RQA indicators $RR$ and $DET$ which works equivalently for regular 
 and chaotic dynamical regimes of both the conservative and the  dissipative
 versions of the employed model approximating an EMRI. Our operational criterion
 allows us to discern deterministic signal from noise up to $ NS \lesssim 40\%$,
 and it remains a robust method even if dissipation is considered. 
 
 Our work is a theoretical dynamical study of a chaotic system with dissipation
 motivated by EMRI systems. This study indicates that chaos is still relevant
 when dissipation and stochastic noise are taken into account.
 In order to explore whether the recurrent analysis can be used as a reliable
 supplementary tool to matched filtering during gravitational data analysis,
 it would require to proceed to more astrophysically relevant signals.
 More relevant types of noise, e.g. frequency dependent, should be employed, and
 the operational criterion should be further fine-tuned to increase
 the critical noise levels. Nevertheless, new templates for matched filtering
 would be necessary in order to obtain values of EMRI's physical parameters
 (as mass and spin), which cannot be deduced from the recurrence analysis.

 \section*{Acknowledgements}
 G.L.-G. and O.K. acknowledge the support from Grant No. GACR-17-06962Y.
 O.K. acknowledges the support from the COST CZ program of the Czech Ministry of
 Education (project LD15061). Discussions with Petra Sukov\'{a} are
 highly appreciated. \\
 
 \appendix
 
 \section{Parameters of the recurrence analysis} \label{sec:RQAparam}
Using the single component approach in the analysis, we reconstruct the
phase space portrait by means of the delay embedding method based on the Takens'
embedding theorems \cite{takens81}. A crucial step is to set the embedding 
dimension ${\cal N}$ corresponding to the dimension of the trajectory's manifold. 
Setting different values of ${\cal N}$ in the analysis we indeed confirm that in our
case the proper value leading to reliable outcome is ${\cal N}=4$
as one would expect in the non-integrable system of two degrees of freedom.
Although the phase space actually has 8 dimensions, only the 4 of them (in our
coordinate choice $\{\rho,~\dot{\rho},~z,~\dot{z}\}$) are dynamically important. 

To construct RPs and compute RQA indicators we use the CRP ToolBox \cite{marwan07}
installed on Matlab (R2014b). The segment of the trajectory we use for the
analysis is given by the range of integration parameter $\tau$ (proper time).
To obtain RPs we use $\tau\in\left<0,2\times10^5\right>$ with the step-size
$\Delta\tau=200$, i.e., 1000 data points. For the evaluation of RQA measures we
use $\tau\in\left<0,4\times10^5\right>$ with $\Delta\tau=200$, i.e., 2000 data
points. We employ Euclidean norm (see Refs.~\refcite{marwan07,kopacek10a} for the
discussion of the choice of the norm in this context). For the recurrence
analysis, the time series is normalized to have zero mean and standard
deviation $\sigma=1$. For the evaluation of the $DET$ indicator we use the 
default value of the minimal length of the diagonal line $l_{\rm min}=2$.

\end{document}